\documentclass[twocolumn,twocolappendix,trackchanges]{aastex7}
\usepackage[version=4]{mhchem}
\usepackage{amsmath}
\usepackage{bm}  
\shorttitle{Coevolution of Dust Grains and Massive Protostellar Disk}
\shortauthors{Yamamuro et al.}
\usepackage{float}
\graphicspath{{./}{figures/}}

\begin{document}

\title{Coevolution of Dust Growth and Gravitational Instability in Massive Protostellar Disks}

\author[0000-0001-7530-1359]{Ryota Yamamuro}
\email{yamamuro.r.aa@m.titech.ac.jp}
\affiliation{Department of Earth and Planetary Sciences, Institute of Science Tokyo, Meguro, Tokyo, 152-8551, Japan}

\author[0009-0006-6613-5751]{Ryoki Matsukoba}
\email{matsukoba@kochi-ct.ac.jp}
\affiliation{National Institute of Technology, Kochi College, 200-1 Monobe, Nankoku, Kochi 783-8508, Japan}

\author[0000-0002-6907-0926]{Kei E. I. Tanaka}
\email{kei.tanaka@eps.sci.titech.ac.jp}
\affiliation{Department of Earth and Planetary Sciences, Institute of Science Tokyo, Meguro, Tokyo, 152-8551, Japan}

\author[0000-0002-6045-0359]{Eduard I. Vorobyov}
\email{eduard.vorobiev@univie.ac.at}
\affiliation{ Institut für Astro- und Teilchenphysik, Universität Innsbruck, Technikerstraße 25, 6020 Innsbruck, Austria}

\begin{abstract}

Massive protostellar disks are expected to sustain high accretion rates through gravitational instability, while dust growth may modify their opacity, cooling efficiency, and stability. Conversely, the time-dependent and nonaxisymmetric structures produced by gravitational instability may regulate the local conditions for dust growth. However, the coupled evolution of dust growth and gravitational instability in massive protostellar disks has not yet been explored self-consistently.
We present two-dimensional hydrodynamic simulations that follow gas dynamics, dust growth, and opacity-dependent radiative cooling in a massive protostellar disk. We find that dust grains rapidly grow to millimeter--centimeter sizes shortly after disk formation, substantially reducing the infrared opacity. The enhanced cooling promotes disk fragmentation and produces more frequent accretion-driven luminosity outbursts than in a model without dust growth, with durations and peak luminosities comparable to those observed in massive protostars. Dust growth is broadly regulated by the fragmentation-limited size in dense disk regions, whereas newly formed outer disk regions and low-density regions deviate from this limit because of finite collision timescales and dust transport. These results suggest that dust growth may facilitate episodic accretion and fragmentation, while gravitationally unstable disk structures in turn regulate the dust growth process.

\end{abstract}

\keywords{Astrophysical dust processes (99); Stellar accretion disks (1570); Star formation (1565); Protostars (1302); Massive stars (732); Protoplanetary disks (1300)}

\section{{Introduction}} \label{sec:intro}

Massive stars exert a profound influence on their surrounding environments through intense radiation of $\gtrsim 10^4\:L_\odot$ and supernova explosions.
Understanding how such massive stars form is therefore a fundamental problem \citep[e.g.,][]{Beuther2025}.
Massive protostars emit strong radiation that inhibits further mass accretion onto the protostar.
As a consequence, it has long been recognized that spherically symmetric accretion from molecular cloud cores imposes an upper limit on the final stellar mass ($\sim 40\:M_\odot$) \citep[e.g.,][]{Larson1971,Kahn1974A, Wolfire1987}.
To mitigate such radiative feedback and enable continued mass supply onto the protostar, the presence of accretion disks around massive protostars has been proposed \citep[e.g.,][]{Yorke2002,Krumholz2009,Kuiper2010,Rosen2016,Tanaka+2017Apj}.
In recent years, high-angular-resolution observations with the Atacama Large Millimeter/submillimeter Array (ALMA) have proposed evidence for the presence of rotationally supported disks around massive protostars \citep[e.g.,][]{Johnstone2015,Cesaroni2017,Maud+2019,Gondi2020,Williams2022,Adam2023,Pan2025,Sanchez-Mong2025,Olguin2026}. However, the internal physical structure and evolution of these disks, including the role of dust growth, remain poorly understood.

In this context, the recent polarimetric observation has suggested that dust grains may grow within massive protostellar disks \citep{Girart2018Ap}.
Dust growth can significantly modify the disk opacity, thereby strongly affecting the disk emission properties at (sub)millimeter wavelengths \citep{Yamamuro2025}.
Such opacity variations introduce systematic uncertainties in disk masses and temperatures inferred from (sub)millimeter observations \citep[e.g.,][]{Vorobyov2026}, highlighting the importance of understanding dust growth processes in massive protostellar disks for the proper interpretation of observations.

Dust growth may also influence the physical evolution of massive protostellar disks themselves. Because these disks are sufficiently optically thick to thermal radiation, their cooling is primarily regulated by dust thermal emission. In this regime, lower opacity allows thermal radiation to escape more efficiently from the disk interior, thereby enhancing the cooling efficiency. Dust growth therefore reduces the infrared opacity and lowers the disk temperature \citep{Yamamuro2023}. This may enhance gravitational instability of the massive protostellar disk.
Since gravitational instability is highly sensitive to the thermal state and cooling efficiency of the disk \citep[e.g.,][]{Gammie2001,Cossins2010,Lee2025}, dust growth may substantially alter the development of gravitational instability and the subsequent evolution of massive protostellar disks.

Such effects are particularly important because massive protostellar disks are thought to maintain their high accretion rates through gravitational instability.
Massive protostellar disks are expected to achieve their high mass accretion rates of $\sim 10^{-4}\:M_\odot\:\mathrm{yr}^{-1}$ via efficient angular momentum transport induced by gravitational instability.
Consequently, nonaxisymmetric spiral structures and disk fragmentation leading to the formation of multiple systems have been suggested from both theoretical and observational perspectives \citep[e.g.,][]{Kratter2010, Beltran2016,Beuther2017,Oliva2020A}.
Furthermore, numerical simulations have demonstrated that the infall of fragments onto the central protostar can trigger episodic accretion bursts in massive protostars \citep[e.g.,][]{Meyer2017,Meyer2018,Elbakyan2023}. Such processes have been proposed as a possible origin of the luminosity outbursts observed in several massive protostars \citep[e.g.,][]{Caratti2017,Zinchenko2017,Hunter2018,Sugiyama2019,Burns2020, Chen2020,Stecklum2021}.

Dust growth in massive protostellar disks has been investigated theoretically by \citet{Yamamuro2023,Yamamuro2025} using steady, axisymmetric disk models. These studies showed that dust grains can grow efficiently in gravitationally unstable massive protostellar disks and that the resulting opacity evolution can modify the disk thermal structure and gravitational stability. However, massive protostellar disks are expected to be time-dependent and nonaxisymmetric, because gravitational instability produces spiral arms, clumps, and episodic accretion. In such disks, dust growth and gravitational instability should influence each other: dust growth changes the opacity and cooling rate of the disk, while the evolving disk structure controls the local density, temperature, and collision timescale that regulate dust growth. It therefore remains unclear whether the dust-growth framework developed for steady disks remains applicable to dynamically evolving gravitationally unstable disks. It is also unclear how dust-growth-induced opacity evolution feeds back on disk fragmentation and accretion variability.

The aim of this study is therefore to investigate how dust growth proceeds in nonsteady and nonaxisymmetric massive protostellar disks and to assess its impact on disk evolution. To this end, we perform self-consistent two-dimensional hydrodynamical simulations that follow the coupled evolution of gas and dust.
This paper is organized as follows. In Section \ref{sec:Method}, we describe the numerical methods and simulation setup. In Section \ref{sec:Result}, we present the main results of the coevolution of the disk, central star, and dust in nonaxisymmetric massive protostellar disks. In Section \ref{sec:CompareDG}, we compare the differences in disk evolution with and without dust growth.
In Section \ref{sec:Dis}, we compare the dust growth in this study with predictions from a 1D steady disk model and show the implications of dust growth for disk gravitational instability.

\section{METHOD} \label{sec:Method}
We simulate gas and dust dynamics in a nonaxisymmetric massive protostellar disk using the numerical hydrodynamics code FEOSAD (Formation and Evolution Of Stars and Disks), which solves the two-fluid (gas and dust) hydrodynamic equations in the thin-disk limit.
The simulations begin with the gravitational collapse of a prestellar cloud core and follow its evolution through central star and disk formation in polar coordinates $(r,\phi)$.
{An important advantage of the two-dimensional thin-disk approach is that it can  capture the effects of gravitational instability and fragmentation, including the formation and evolution of spiral arms and clumps, while allowing for substantially higher numerical resolution at a considerably lower computational cost than full three-dimensional simulations. This makes the 2D thin-disk approximation particularly suitable for following the long-term dynamical evolution of massive protostellar disks and the associated dust evolution.}
A detailed description of the code is provided in \citet{Vorobyov+2018, Vorobyov2023}. Here, we briefly summarize its key features and describe the updates implemented for the present study.

\subsection{Basic Equations for Gaseous Component}

The FEOSAD code considers the coevolution of gas and dust disk systems, treating them as two fluids, with the dust represented as a pressureless fluid.
The hydrodynamic equations for the gas, describing the conservation of mass and momentum, are given by
\begin{equation}
    \frac{\partial \Sigma_\mathrm{g}}{\partial t} + \nabla \cdot (\Sigma_\mathrm{g} \bm{v})=0,
\end{equation}

\begin{equation}
    \frac{\partial}{\partial t} (\Sigma_\mathrm{g} \bm{v}) + \nabla \cdot (\Sigma_\mathrm{g}\bm{v}\: \otimes  \bm{v} )= -\nabla{\mathcal{P}} +\Sigma_\mathrm{g}\bm{g}\:+ \nabla\cdot \bm{\Pi} -\Sigma_\mathrm{d,\:gr}  \bm{f},
\end{equation}
where $\Sigma_\mathrm{g}$ is the gas surface density, $\Sigma_\mathrm{d,gr}$ is the surface density of the grown dust component, and $\bm{f}$ is the gas–dust drag force per unit grown-dust mass described in next section, while $\nabla=\hat{r},\partial/\partial r + \hat{\phi},r^{-1}\partial/\partial\phi$ denotes the planar gradient and $\bm{v}=v_r\hat{r}+v_\phi\hat{\phi}$ is the planar gas velocity. The vertically integrated pressure is given by $\mathcal{P}=(\gamma-1)e$ with $\gamma=7/5$ and internal energy per unit surface area $e$.
The gravitational acceleration $\bm{g}=g_r\hat{r}+g_\phi\hat{\phi}$ includes the self-gravity of the gas and dust disk components obtained from the integral form of the Poisson equation \citep{Vorobyov2024} together with the gravity of the central protostar once it forms. The viscous stress tensor $\mathbf{\Pi}$ follows the formulation of \citet{Vorobyov2010}.
We parameterize the magnitude of kinematic viscosity $\nu=\alpha c_\mathrm{s} H_\mathrm{g}$ using the alpha prescription of \cite{Shakura1973}, where $c_\mathrm{s}$ is the sound speed of gas and $H_\mathrm{g}$ is the gas vertical scale height \citep{Vorobyov2009}.
The $\alpha$ is set to a spatially and temporally constant value.
{Massive protostellar disks are expected to exhibit strong turbulence in addition to gravitational torques in order to sustain their high mass accretion rates. We therefore adopt $\alpha = 10^{-2}$, following the assumption of previous study \citep{Yamamuro2023}.}

The energy balance equation, including viscous accretion heating and compressional heating, is written as
\begin{equation}
    \frac{\partial e}{\partial t} +\nabla\cdot(e\bm{v})=-\mathcal{P}(\nabla\cdot \bm{v})-\Lambda+\Gamma +(\nabla\bm{v}):\Pi,
\end{equation}
where $\Lambda$ and $\Gamma$ denote the cooling and heating terms, respectively.
These terms account for disk blackbody cooling and for heating due to stellar and background irradiation, following the analytical solution to the vertical radiative transfer equation by \citet{Dong2016,Pavlyuchenkov2023}.
The cooling term is given by
\begin{equation}
      \label{eq:cooling}
    \Lambda=\frac{8 \sigma_\mathrm{SB} T^4 }{2 + \frac{3}{2}\tau_\mathrm{R} +\frac{1}{\tau_\mathrm{P}}  },
\end{equation}
where $T=(\gamma-1)e \mu/\mathcal{R}\Sigma_\mathrm{gas} $ is the midplane temperature, $\mu = 2.33$ is the mean molecular weight, $\mathcal{R}$ is the universal gas constant, and $\sigma_\mathrm{SB}$ is the Stefan–Boltzmann constant.
The Rosseland and Planck optical depths to the disk midplane are defined as $\tau_\mathrm{R}=\kappa_\mathrm{R} \Sigma_\mathrm{d, tot}/2$ and $\tau_\mathrm{P}=\kappa_\mathrm{P}\Sigma_\mathrm{d, tot}/2$, where $\Sigma_\mathrm{d, tot}$ is the total dust surface density. 
The corresponding mean opacities $\kappa_\mathrm{R}$ and $\kappa_\mathrm{P}$ depend on the maximum grain radius $a_\mathrm{max}$ (see Section \ref{sec:Dustop}).
The heating term per unit surface area of the disk is expressed as
\begin{equation}
    \Gamma=\frac{8 \sigma_\mathrm{SB} T_\mathrm{irr}^4 }{2+ \frac{3}{2}\tau_\mathrm{R}  +\frac{1}{\tau_\mathrm{P}}  },
\end{equation}
where $T_\mathrm{irr}$ is the irradiation temperature at the disk surface determined by the stellar and background irradiation as
\begin{equation}
    T^4_\mathrm{irr}=T^4_\mathrm{bg}+\frac{F_\mathrm{irr}}{\sigma_\mathrm{SB}},
\end{equation}
where $T_\mathrm{bg}$ is background temperature set to $10\:\mathrm{K}$, $F_\mathrm{irr} (r)$ is the radiation flux absorbed by the disk surface at radial distance $r$, given by
\begin{equation}
    F_\mathrm{irr} (r)= \frac{L_*}{4\pi r^2}\mathrm{cos} \:{\gamma_\mathrm{irr}},
\end{equation}
where $\gamma_\mathrm{irr}$ is the incidence angle at $r$ described in \cite{Vorobyov2010}.
The stellar luminosity $L_*$ is the sum of the accretion luminosity and the stellar intrinsic luminosity.
We evaluate the stellar intrinsic luminosity from pre-calculated stellar evolutionary tracks derived from \cite{Hosokawa2009} as a function of the stellar mass and accretion rate.
Here, we use the time-averaged accretion rates for the previous $1\:\mathrm{yr}$ because it fluctuates significantly in each time steps.

\subsection{Basic Equations for Dusty Components}
The dust in our model consists of two components. The small dust component has grain radius between $a_\mathrm{min} = 5\times10^{-3}\:\mathrm{\mu m}$ and $a_* = 1\:\mathrm{\mu m}$, while the grown dust component has radius from $a_*$ up to a maximum grain radius $a_\mathrm{max}$, which varies in space and time.
Initially, the prestellar core contains only small grains with $a_\mathrm{max} = a_*$.
As the disk forms and evolves, small grains can grow and transition into the grown dust component.
Both small and grown dust follow a power-law size distribution,
\begin{equation}
\label{Eq:Na}
    N(a)=C\cdot a^{-q},
\end{equation}
{where $N(a)$ is the surface number density of grains per unit grain radius $a$ and $q=3.5$ is the power-law index of the dust-size distribution. Following the classical theory of fragmentation cascades \citep{Dohnanyi1969,Birnstiel2011}, we adopt $q=3.5$ for simplicity.
$C$ is a normalization constant determined by
\begin{equation}
\int_{a_\mathrm{min}}^{a_\mathrm{max}} m_\mathrm{d}(a) \cdot N(a) da
= \Sigma_\mathrm{d,sm} + \Sigma_\mathrm{d,gr},
\end{equation}
where $m_\mathrm{d}(a)$ is the mass of an individual grain with radius $a$ and $\Sigma_\mathrm{d,sm}$ is the surface density of small dust.}

The continuity and momentum conservation equations for small and grown dust are given by
\begin{equation}
\frac{\partial \Sigma_\mathrm{d,sm}}{\partial t} + \nabla \cdot (\Sigma_\mathrm{d,sm} \bm{v}) = -S(a_\mathrm{max}),
\end{equation}
\begin{equation}
\frac{\partial \Sigma_\mathrm{d,gr}}{\partial t} + \nabla \cdot (\Sigma_\mathrm{d,gr} \bm{u}) = S(a_\mathrm{max}),
\end{equation}
\begin{equation}
\frac{\partial}{\partial t} (\Sigma_\mathrm{d,gr} \bm{u}) + \nabla \cdot (\Sigma_\mathrm{d,gr} \bm{u} \otimes \bm{u}) = \Sigma_\mathrm{d,gr} \bm{g} + \Sigma_\mathrm{d,gr} \bm{f} + S(a_\mathrm{max}) \bm{v},
\end{equation}
where $\Sigma_\mathrm{d,sm}$ is assumed to be perfectly coupled to the gas and thus shares the gas velocity $\bm{v}$, while $\bm{u} = u_r \hat{r} + u_\phi \hat{\phi}$ is the planar velocity of grown dust. 
The term $\bm{f}$ represents the drag force per unit mass of grown dust due to interaction with the gas and is defined as
\begin{equation}
    \bm{f}=\frac{1}{2m_\mathrm{d}} C_{D} \sigma \rho_\mathrm{g} (\bm{v} - \bm{u}) |\bm{v} - \bm{u}|,
\end{equation}
where $\sigma = \pi a_\mathrm{max}^2$ is the cross-section of dust grains, $\rho_\mathrm{g}$ is the gas volume density, and $C_{D}$ is the dimensionless friction coefficient.
The functional form of $C_{D}$ is taken from \citet{Henderson1976} and is described in detail in \citet{Stoyanov2020} and \citet{Vorobyov2023}.
Using the Henderson friction coefficient allows the drag force to be treated self-consistently in both the Epstein and Stokes regimes, reproducing effects such as the sonic boom and avoiding discontinuities, which may occur in the more commonly used drag coefficient of \citet{Weidenschilling1977}.
The stopping time associated with the Henderson drag is defined as
\begin{equation}
  t_\mathrm{stop} = \frac{8}{3}\,\frac{a\,\rho_{\mathrm{s}}}{\rho_{\mathrm{g}}\,C_D\,\lvert\bm{v}-\bm{u}\rvert}\ ,
  \label{Eq;Henderson-stop-time}
\end{equation}
where $\rho_\mathrm{s} = 1.68\:\mathrm{g\:cm^{-3}}$ is the internal density of dust grains \citep{Birnstiel2018}. In calculating both the drag coefficient and the stopping time, we adopt $a_\mathrm{max}$ as the characteristic grain radius.
Since the grain-size distribution follows a power law with $q = 3.5$, the dust surface density is dominated by the largest grains \citep{Sato+2016}. Therefore, we track the dynamics of the grains that carry most of the dust mass.
The term $S(a_\mathrm{max})$ in the dust equations represents the conversion rate between the small and grown dust components.
The dust grain distribution for both components is assumed to follow the power-law form given in Equation (\ref{Eq:Na}), and the grain-size distribution is taken to be continuous at $a_*$.
Our scheme ensures this continuity by defining the conversion rate from small to grown dust accordingly (see \citealt{Vorobyov2023}).
The conversion rate $S(a_\mathrm{max})$ depends only on the local maximum grain radius $a_\mathrm{max}$, as $a_\mathrm{min}$ and $a_*$ are fixed in our model.

At the beginning of the prestellar cloud core collapse, all grains belong to the small dust component with $a_\mathrm{max} = a_*$.
During the subsequent disk formation and evolution, the high-density disk environment facilitates frequent dust–dust collisions, leading to an increase in the maximum grain radius $a_\mathrm{max}$.
In our model, the evolution of $a_\mathrm{max}$ within a numerical cell occurs due to collisional growth or fragmentation, as well as advection of dust through the cell.
The dynamical evolution of $a_\mathrm{max}$ is described by
\begin{equation}
\label{Eq:dustgrowth}
    \frac{\partial a_\mathrm{max}}{\partial t} + (\bm{u}\cdot\nabla)a_\mathrm{max}=\xi_\mathrm{frag} \mathcal{D},
\end{equation}
where the collisional growth rate $\mathcal{D}$ is computed in the monodisperse approximation \citep{Binrstiel2012},
\begin{equation}
    \mathcal{D}=\frac{\rho_\mathrm{d} u_\mathrm{rel}}{\rho_\mathrm{s}},
\end{equation}
here, $\rho_\mathrm{d} = \Sigma_\mathrm{d,tot} / \sqrt{2\pi}H_\mathrm{d}$ is the total dust volume density, where the dust scale height $H_\mathrm{d}$ is taken into account through dust settling \citep{Kornet2001}. The relative collision velocity is defined as $u_\mathrm{rel} = \sqrt{u_\mathrm{turb}^2 + u_\mathrm{th}^2}$, where $u_\mathrm{turb}$ and $u_\mathrm{th}$ represent the turbulence-induced and Brownian-motion components, respectively.
The collisional velocity induced by turbulence is derived using the analytical expression of \citet{Ormel2007}, 
\begin{equation}
    u_\mathrm{turb}=\sqrt{\frac{3\alpha {c_s}^2}{\mathrm{St}+\mathrm{St}^{-1}}},
\end{equation}
where $\mathrm{St}$ is the Stokes number, defined as
\begin{equation}
\label{eq:St}
   \mathrm{St}=t_\mathrm{stop} \Omega_\mathrm{K},
\end{equation}
where $\Omega_\mathrm{K}$ is the Keplerian angular velocity {and the stopping time $t_{\rm stop}$ is calculated for the representative grain radius $a_{\rm max}$, see Equation~(\ref{Eq;Henderson-stop-time}). }
{Therefore, the relative collision velocity due to turbulence $u_{\rm turb}$ is also evaluated for grains of representative radius $a_\mathrm{max}$. }
The thermal velocity due to the Brownian motion is calculated as
\begin{equation}
    u_\mathrm{th}=\sqrt{\frac{16k_\mathrm{B} T}{\pi m_\mathrm{d}}},
\end{equation}
where $k_\mathrm{B}$ is the Boltzmann constant {and the mass of dust grain $m_{\rm d}$ is also calculated for the represnetative grain radius $a_{\rm max}$. We note that $u_{\rm th}$ is usually much smaller than $u_{\rm turb}$ and dust growth due to the Brownian motion becomes important only for $a_{\rm max}\simeq a_\ast$. }

Next, we adopt the following form for the fragmentation term,
\begin{equation}
   \xi_\mathrm{frag}=    \mathrm{min} \left(1,-\frac{\mathrm{ln}(u_\mathrm{rel}/u_\mathrm{frag})}{\mathrm{ln}5}\right), 
\end{equation}
where $u_\mathrm{frag}$ is the fragmentation velocity, which we set $u_\mathrm{frag} = 10\:\mathrm{m\:s^{-1}}$ in this study, following \citet{Yamamuro2023,Yamamuro2025}. If the collisional velocity exceeds this threshold, net erosion of the largest grains occurs \citep{Okuzumi2016}, leading to a decrease in $a_\mathrm{max}$.
In the conventional FEOSAD framework, when the relative velocity of dust grains exceeds the fragmentation threshold, the maximum grain radius is instantaneously reset to the fragmentation-limited value \citep[e.g.,][]{Vorobyov+2018}.
However, collisional fragmentation is fundamentally a process that occurs only when actual collisions between dust grains take place.
Using Equation (\ref{Eq:dustgrowth}), the typical timescale for such collisions can be expressed as
\begin{equation}
    t_\mathrm{coll}
    = \frac{a_\mathrm{max}}{\mathcal{D}}
    = \frac{\rho_\mathrm{s} a_\mathrm{max}}{\rho_\mathrm{d} u_\mathrm{rel}}.
\end{equation}
Since this collision timescale often exceeds the local Keplerian timescale \citep{Brauer2008,Sato+2016}, we introduce a source term, $\xi_\mathrm{frag}$, to account for mass loss due to fragmentation in a manner consistent with the finite collision timescale. {We also note that dust fragmentation in our simplified model replenishes dust populations of all sizes down to $a_{\rm min}$, while keeping the power index fixed.}

\subsection{Dust Opacity} 
\label{sec:Dustop}
In circumstellar disk environments, dust grains can grow by several orders of magnitude from their initial interstellar sizes of approximately micrometers \citep{Weingartner2001}. Such rapid growth reduces the dust opacity at infrared wavelengths. As a result, the cooling efficiency of the disk increases, leading to a lower disk temperature.

To account for these changes, we employ the publicly available code {\tt Optool} \citep{Dominik2021} to compute the Rosseland and Planck mean opacities of dust.
This code generates opacities based on the input parameters of dust structure, size distribution, and composition. In this study, we assume that the dust grains are spherical and compact, and the dust-size distribution follows $q=3.5$.
The dust opacities per unit dust mass are then computed accordingly. The dust composition is taken to be the ice-rich mixture adopted in the DSHARP project \citep{Birnstiel2018}.
Using this tool, we obtain the Rosseland mean opacity 
\begin{equation}
   \cfrac{1}{\kappa_\mathrm{R}(a_\mathrm{max},T)}=
\cfrac{\displaystyle\int^\infty_0 (\kappa_{\nu, \mathrm{abs} } + \kappa^\mathrm{eff}_{\nu,\mathrm {sca} })^{-1} \cfrac{\partial B_\nu(T)}{\partial T} d \nu}
{\displaystyle\int^\infty_0 \cfrac{\partial B_\nu(T)}{\partial T} d \nu},
\end{equation}
and the Planck mean opacity
\begin{equation}
    \kappa_\mathrm{P} (a_\mathrm{max}, T)= \cfrac{\displaystyle\int^\infty_0 \kappa_{\nu,\mathrm{abs}}  B_\nu(T)}
{\displaystyle\int^\infty_0  B_\nu(T) d \nu},
\end{equation}
where $\kappa_{\nu, \mathrm{abs}}$, $\kappa^{\mathrm{eff}}_{\nu, \mathrm{sca}}$, and $B_\nu$ are the monochromatic absorption and effective scattering opacities at frequency $\nu$, and the Planck function, respectively.

\begin{figure}%[H]
\includegraphics[width=\hsize]{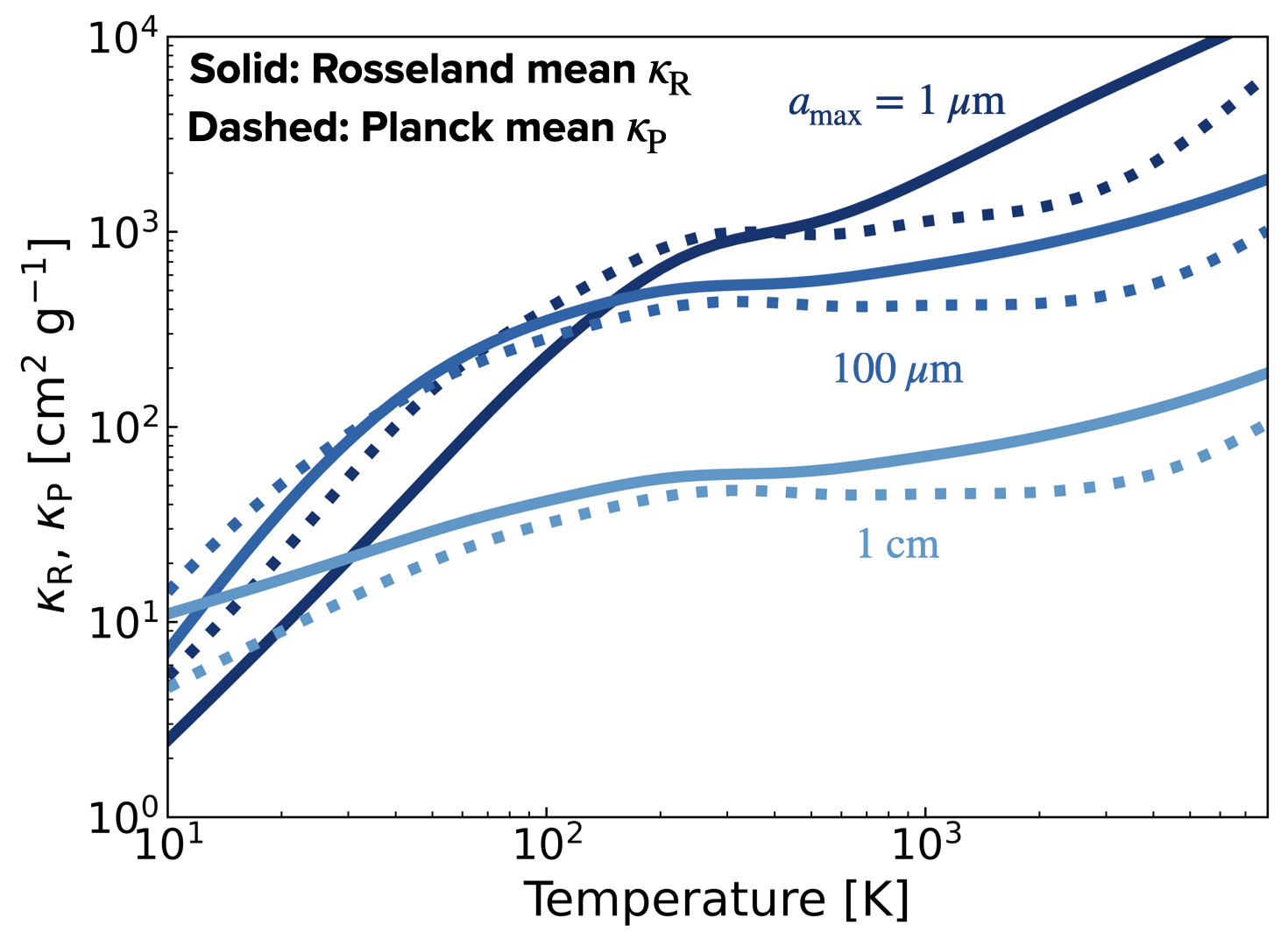}\
\caption{The Rosseland mean opacity $\kappa_\mathrm{R}$ (solid lines) and Planck mean opacity $\kappa_\mathrm{P}$ (dashed lines) per dust mass as a function of temperature at the maximum grain radius of $1\:\mathrm{\mu m}, 100\:\mathrm{\mu m},$ and $1\:\mathrm{cm}$ (deep blue, blue, and light blue lines respectively).  }
\label{DustOpa}
\end{figure}

{To demonstrate the effect of dust growth on opacity, we show in Figure \ref{DustOpa} the Rosseland and Planck mean opacities as a function of temperature for maximum grain radii of $1\:\mathrm{\mu m}$, $100\:\mathrm{\mu m}$, and $1\:\mathrm{cm}$. Both the Rosseland and Planck mean opacities generally increase with temperature.
While smaller grains typically produce higher opacity, this trend reverses over certain temperature ranges.
This reversal is related to the wavelength at which the Planck function $B_\nu$ peaks for a given temperature $T$, corresponding to the wavelength at which the thermal emission peaks, $\lambda_\mathrm{peak} = 10\:(300\:\mathrm{K}/T)\:\mathrm{\mu m}$, relative to the maximum grain radius.
When $\lambda_\mathrm{peak} < 2\pi a_\mathrm{max}$, grains enter the efficient absorption and scattering regime at the wavelengths dominating the thermal emission.}

As a result, in the temperature range
$10$--$100\:\mathrm{K}$ ($\lambda_\mathrm{peak}\sim30$--$300\:\mathrm{\mu m}$), dust with
$a_\mathrm{max}=100\:\mathrm{\mu m}$ or $1\:\mathrm{cm}$ exhibits higher opacity than dust with $a_\mathrm{max}=1\:\mathrm{\mu m}$.
However, at temperatures above $\sim100\:\mathrm{K}$, which characterize most of the disk regions analyzed in this study, the opacity decreases with increasing grain size.
In this regime, even relatively small grains efficiently absorb and scatter radiation at the wavelengths contributing most to the mean opacity, and larger grains provide a smaller opacity per unit mass.

In the temperature range $300$--$1000\:\mathrm{K}$, the opacity increases more gradually for all grain radius, which can be attributed to the silicate feature around $10\:\mathrm{\mu m}$ included in the adopted water-ice dust mixture (e.g., \citealt{Draine2003}).
At temperatures above $\sim1000\:\mathrm{K}$, $\lambda_\mathrm{peak}$ shifts to wavelengths shorter than $\sim3\:\mathrm{\mu m}$.
Consequently, the peak of the thermal emission shifts into a wavelength regime where the dust opacity rises toward shorter wavelengths. As a result, the Rosseland mean opacity increases, and this trend dominates over the opacity structure associated with the $10\:\mathrm{\mu m}$ silicate feature.

Note that massive protostellar disks are sufficiently warm that ice-free dust would in principle be more appropriate in the innermost regions. However, the ice-free region is expected to be limited to radii of order $\sim100\:\mathrm{au}$, whereas our computational domain extends to $\sim0.1\:\mathrm{pc}$. Moreover, ice-free dust exhibits strong silicate features that introduce a sharp opacity increase around $10\:\mathrm{\mu m}$. To isolate the impact of dust-growth-driven opacity variations on disk evolution, we therefore adopt water-ice dust throughout the calculation. Consistent with this simplification, we also neglect additional processes such as the evaporation of volatile species, complete dust sublimation, and changes in the gas opacity caused by molecular dissociation and ionization.

\subsection{Initial and Boundary Conditions}
\label{subsec:InitialCondition}
We start the simulation from a pre-stellar cloud core and make it gravitationally unstable to trigger collapse.
The initial cloud profiles have a plateau with axisymmetric uniform gas surface density and angular velocity \citep{Vorobyov2010}
\begin{equation}
    \Sigma_\mathrm{g}=\frac{\Sigma_0}{\sqrt{1+ (r/r_0)^2}},
\end{equation}
\begin{equation}
    \Omega=2\Omega_0 \left(\frac{r_0}{r}\right)^2 \left[\sqrt{1+ \left(\frac{r}{r_0} \right)^2} -1 \right],
\end{equation}
where $\Sigma_0$ is the gas surface density at the plateau, $\Omega_0$ is the angular velocity at the plateau, and $r_0$ is the plateau radius.
This profile emerges when an axisymmetric cloud collapses under its own gravity while conserving angular momentum \citep{Basu1997}.
The plateau radius $r_0$ is proportional to the Jeans length
\begin{equation}
    r_0=A\frac{c_{s,0}}{\sqrt{\pi G \rho_0}},
\end{equation}
where $A$ is a constant parameter, $c_{s,0}$ is the initial sound speed, $G$ is the gravitational constant, and $\rho_0$ is the initial mass density.
The initial cloud has a total mass of $M_\mathrm{cloud} = 30\:M_\odot$ with a plateau radius $r_0 = 3.9\times10^3\:\mathrm{au}$ and surface density $\Sigma_0 = 0.4\:\mathrm{g\:cm^{-2}}$.
The surface density is set such that the mass infall rate onto the central star plus disk system, estimated from $M_\mathrm{core}/t_\mathrm{ff}$ with the free-fall time $t_\mathrm{ff} = \sqrt{3\pi / 32 G \rho_0} \approx 25\:\mathrm{kyr}$, is $\sim 5\times10^{-4}\:M_\odot\:\mathrm{yr}^{-1}$ , which is comparable to typical accretion rates inferred for massive protostars \citep{Mckee2002,Hosokawa2009}.
The ratio of rotational to gravitational energies is $0.4\:\%$, the rotational velocity at the plateau radius is $r_0\Omega_0 = 2.2\:\mathrm{km\:s^{-1}}$, and the initial temperature is $T_\mathrm{bg} = 10\:\mathrm{K}$.
All dust is initially in the small dust component with $a_\mathrm{max} = a_*$, and the dust-to-gas mass ratio is set to $\xi_\mathrm{d2g} = 0.01$.

{The hydrodynamic equations are solved on the polar grid ($r, \phi$) using the method of finite volumes described in the ZEUS code \citep{StoneNorman1992}. The hydrodynamic quantities are advected using a third-order piecewise parabolic scheme of \citet{Colella1984}, ensuring low numerical diffusion. The computational domain extends from 5 au to $3\times 10^4\:$au and the number of grid cells is $512 \times 512$.  The grid is logarithmically spaced in the radial direction and equidistantly spaced along the azimuthal coordinate, which corresponds to a numerical resolution of $0.1$ au near the inner computational boundary and $1.7\:\mathrm{au}$ at $r=100$ au.
To quantitatively justify the spatial resolution, we define the Jeans length estimated from the azimuthally averaged surface density and sound speed as $R_j = c^2_{s,\mathrm{ave}} / \pi G \Sigma_{\mathrm{g,ave}}$ \citep{Vorobyov_2013}. With this definition, the spatial resolution satisfies the Truelove criterion, resolving the Jeans length by more than four grid cells throughout the disk in the immediate vicinity of dense clumps at both $t=40$ and $150\:\mathrm{kyr}$ in Section \ref{sec:Result} \citep{Truelove1998}.

As the cloud core contracts gravitationally, it spins up and a centrifugal disk forms when the in-spiralling gas and dust hit the centrifugal barrier near the inner computational boundary.
Subsequently, the disk grows in size and mass owing to infall from the cloud, which occurs at the disk's outer edge in our model, typical of flared disks \citep[see][]{Visser2009}, while the central star gains mass via accretion through the inner computational boundary.
The latter is placed at $5\:\mathrm{au}$ and the regions interior to $5\:\mathrm{au}$ are replaced with a sink cell. Instead of the standard one-way open boundary condition at the inner boundary, which allows matter to flow through the disk-sink interface toward the star but prohibits the flow in the opposite direction, we employ a more sophisticated scheme, in which matter is allowed to travel both ways across the sink-disk interface. The details of this boundary condition can be found in \citet{Vorobyov+2018}, with comparison against the one-way boundary condition in \citet{Matsukoba2025}. Our updated boundary condition provides a compensating flow of matter from the sink to the disk and mitigates the problem of a sharp drop in density that may occur when the one-way boundary condition is applied.
{This bidirectional flow does not violate the total mass conservation of the star-disk-envelope system. It also has a physical motivation, as it may partially account for local outward mass transport associated with angular-momentum redistribution in gravitationally unstable disks during disk accretion, which would otherwise be artificially suppressed by a one-way sink boundary.}
The outer boundary condition is the standard one-way outflow, in which matter is allowed to flow out of the computational domain but not allowed to flow in. Since the outer boundary is placed far away from the disk, it has no effect on the disk dynamics and merely serves to prevent matter from external environment to flow into the computational domain.}

\section{Evolution of a nonaxisymmetric massive protostellar disk with dust growth} \label{sec:Result}

In this section, we present the results of disk evolution and dust growth in a massive protostellar disk.
We first show the spatial distributions of the main disk quantities in Section \ref{subsec:Spa}. We then describe the time evolution of the stellar and disk properties in Section \ref{subsec:Global}.
In Section \ref{subsec:gas}, we present the radial distributions of the main gas properties for a quantitative evaluation of disk evolution. Finally, Section \ref{sec:Dust} provides a quantitative analysis of dust growth.

\label{subsec:Ref}

\subsection{Snapshots of Disk Evolution}
\label{subsec:Spa}
\begin{figure}
\includegraphics[width=\hsize]{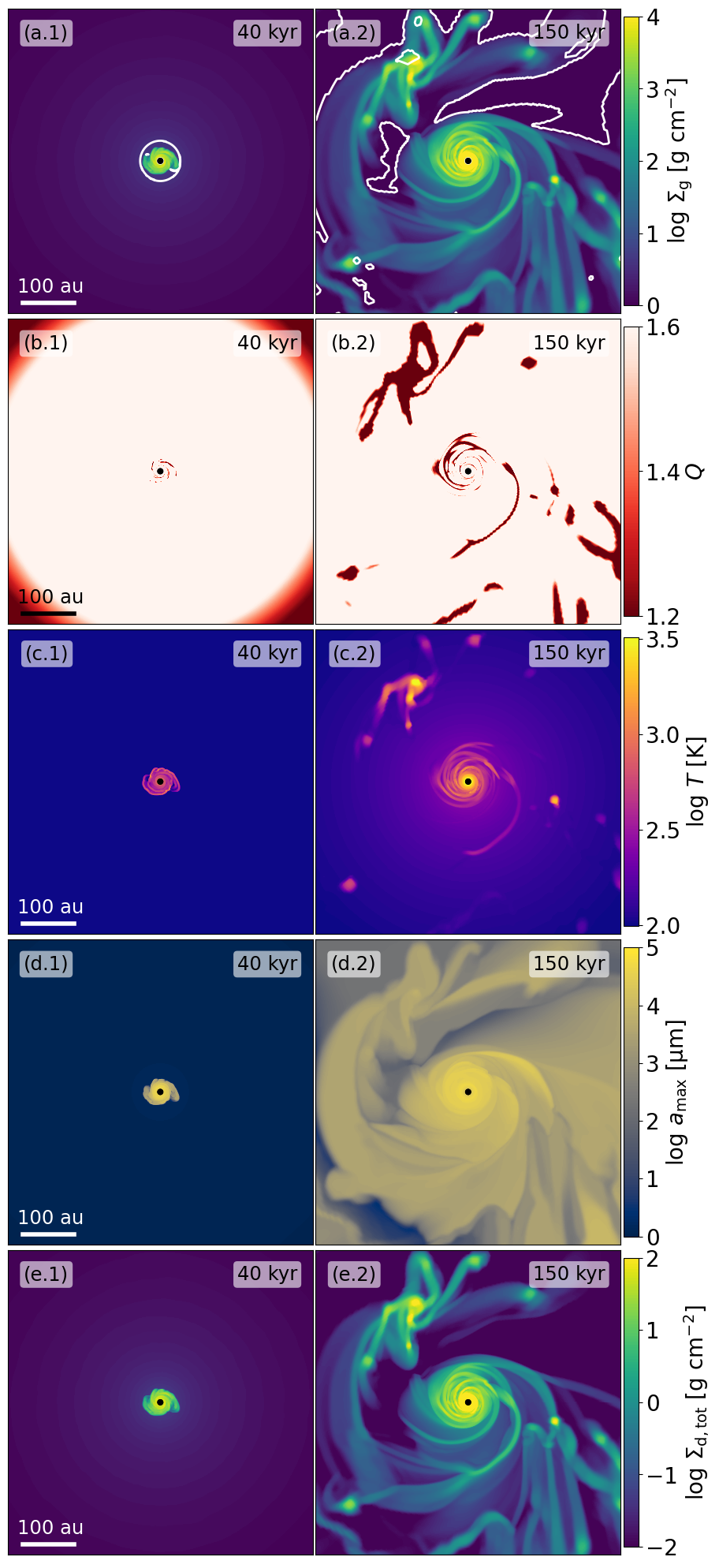}\
\caption{Snapshots of the gas surface density, Toomre's $Q$, temperature, maximum grain radius, tand total dust surface density at $t = 40$ and $150\:\mathrm{kyr}$, in the inner $600\times600 \:\mathrm{au}^2$ box for the model with dust growth.
The white contours in panels (a.1) and (a.2) indicate the boundary of the disk region.}
\label{Gro}
\end{figure}

We present the spatial distributions of the main disk quantities to provide an overview of the disk evolution.
Figure \ref{Gro} shows the snapshots of the primary physical quantities within the inner $600\times600\:\mathrm{au}$ region at two epochs $t = 40$ and $150\:\mathrm{kyr}$. These snapshots represent different evolutionary stages.
The former corresponds to the early disk-formation phase, whereas the latter corresponds to a later stage of the main accretion phase, by which point the central star has already become a massive protostar, reaching roughly half of the total system mass ($16\:M_\odot$; see the solid blue line in panel (a) of Figure \ref{Star_Fig_Dust}).

We define the disk region based on the centrifugal radius. At a distance $r$, the centrifugal radius is calculated from the gas azimuthal velocity $v_\phi$ and the total enclosed mass within $r$, which includes the central star and the gas mass, $M_\mathrm{star}+M_\mathrm{cum}(r)$, as
\begin{equation}
\label{eq:rdisk}
r_\mathrm{cent} = \frac{G (M_\mathrm{star}+M_\mathrm{cum}(r))}{{v_\phi}^2}.
\end{equation}
For each azimuthal angle on the computational grid, the disk region is identified as the contiguous set of cells extending outward from the sink cell that satisfy the condition $r_\mathrm{cent}/r < \gamma_\mathrm{cent}$ \citep{Kadam2025}. We adopt a threshold value of $\gamma_\mathrm{cent} = 1.5$ throughout this study. To further constrain the disk radius, we impose a lower limit on the gas surface density, requiring $\Sigma_\mathrm{g} > \Sigma_0 = 0.4\:\mathrm{g\:cm^{-2}}$, corresponding to the initial molecular cloud core.

Panel (a.1) of Figure \ref{Gro} shows the gas surface density at $t=40\:\mathrm{kyr}$, shortly after disk formation at $t=37\:\mathrm{kyr}$. At this epoch, spiral arms are already formed, and the disk radius is $\sim 40\:\mathrm{au}$.
The disk radius is indicated by the white contour, which corresponds to a threshold density. 
As accretion from the envelope continues, the disk expands.
Panel (a.2) illustrates the development of nonaxisymmetric structures, such as spiral arms and clumps. At this stage, the disk radius reaches $\sim 300\:\mathrm{au}$, although the presence of gap regions associated with disk fragmentation leads to strong azimuthal variations, with the disk radius decreasing to $\sim 100\:\mathrm{au}$ in some directions.

To investigate the origin of these structures, we examine Toomre's $Q$ parameter \citep{Toomre1964}, defined as
\begin{equation}
Q=\frac{\tilde{c}_s \Omega}{\pi G (\Sigma_\mathrm{g} + \Sigma_\mathrm{d,tot})},
\end{equation}
where $\tilde{c}_s = c_s/\sqrt{1+\xi_\mathrm{d2g}}$ is the modified sound speed \citep{Vorobyov+2018} in the presence of dust, $\Omega$ is the angular velocity of gas.
As shown in Panels (b.1)–(b.2), the spiral-arm and clump regions correspond to $Q \lesssim 1.4$, indicating that these structures arise from the gravitational instability \citep{Durisen2007}.

Panels (c.1)–(c.2) show the disk temperature structure. 
The heating sources are radiative heating from the central star, viscous accretion heating, and compressional heating due to disk self-gravity.
At both epochs, temperature enhancements are found within spiral arms at radii $\lesssim 50\:\mathrm{au}$. At $t = 150\:\mathrm{kyr}$, additional localized enhancements appear within clumps, where the temperature reaches up to $\sim1000\:\mathrm{K}$.
These regions coincide with the spiral arms and clumps shown in panels (a.1) and (a.2), indicating that the temperature increase is caused by viscous accretion heating and, mainly, by compressional heating associated with self-gravity.
As the stellar mass and luminosity increase (solid blue lines in panels (a) and (c) of Figure \ref{Star_Fig_Dust}), radiative heating strengthens, leading to a corresponding increase in temperature from panels (c.1) to (c.2). Based on the increase in luminosity, the temperature is expected to rise by more than a factor of $\sim 3$ in regions where radiative heating dominates.

Panels (d.1)–(d.2) show the distribution of the maximum grain radius. 
The maximum grain radius reaches $\sim 10^3$--$10^4\:\mathrm{\mu m}$ within $\sim 20\:\mathrm{au}$ at the early disk-formation stage ($t = 40\:\mathrm{kyr}$).
As the disk expands, the region where grains grow gradually widens, and eventually grains reach radii of $\sim 10^4\:\mathrm{\mu m}$ throughout most of the inner $300\:\mathrm{au}$.
This is related to efficient collisional dust growth in dense spiral arms and clumps, from which grown dust grains leak out into the surrounding low-density regions (see Section \ref{sec:Dust} for details).

Panels (e.1)–(e.2) display the total dust surface density structure.
The dust surface density generally follows the behavior of the gas disk over most regions. Although dust growth could, in principle, produce structures arising from dust–gas decoupling \citep{Whipple1972,Adachi1976, Vorobyov+2018,Matsukoba2025}, no such features appear in our calculations. 
Within $\lesssim 40\:\mathrm{au}$ at $t=40\:\mathrm{kyr}$ and in both the inner disk ($\lesssim 60\:\mathrm{au}$) and the outer dense spiral-arm and clump regions ($\gtrsim 60\:\mathrm{au}$) at $t=150\:\mathrm{kyr}$, the Stokes number remains $\mathrm{St}\lesssim10^{-3}$, indicating tight dust-gas coupling.
In the low-density regions beyond $\gtrsim 60\:\mathrm{au}$ at $t=150\:\mathrm{kyr}$, however, grown dust leaks out of the dense spiral arms and clumps, as shown in panel (d.2). Since the Stokes number is inversely proportional to the local gas surface density (Equation \ref{eq:St}), $\mathrm{St}$ approaches unity in these regions, leading to partial dust--gas decoupling and dust depletion. Although dust can decouple from the gas in the low-density regions, the dust density there is sufficiently low that the resulting dust drift has little effect on the dust-to-gas ratio in the nearby spiral arms and clumps.

\subsection{Time Evolution of the Protostar–Disk System}
\label{subsec:Global}

\begin{figure}
\includegraphics[width=\hsize]{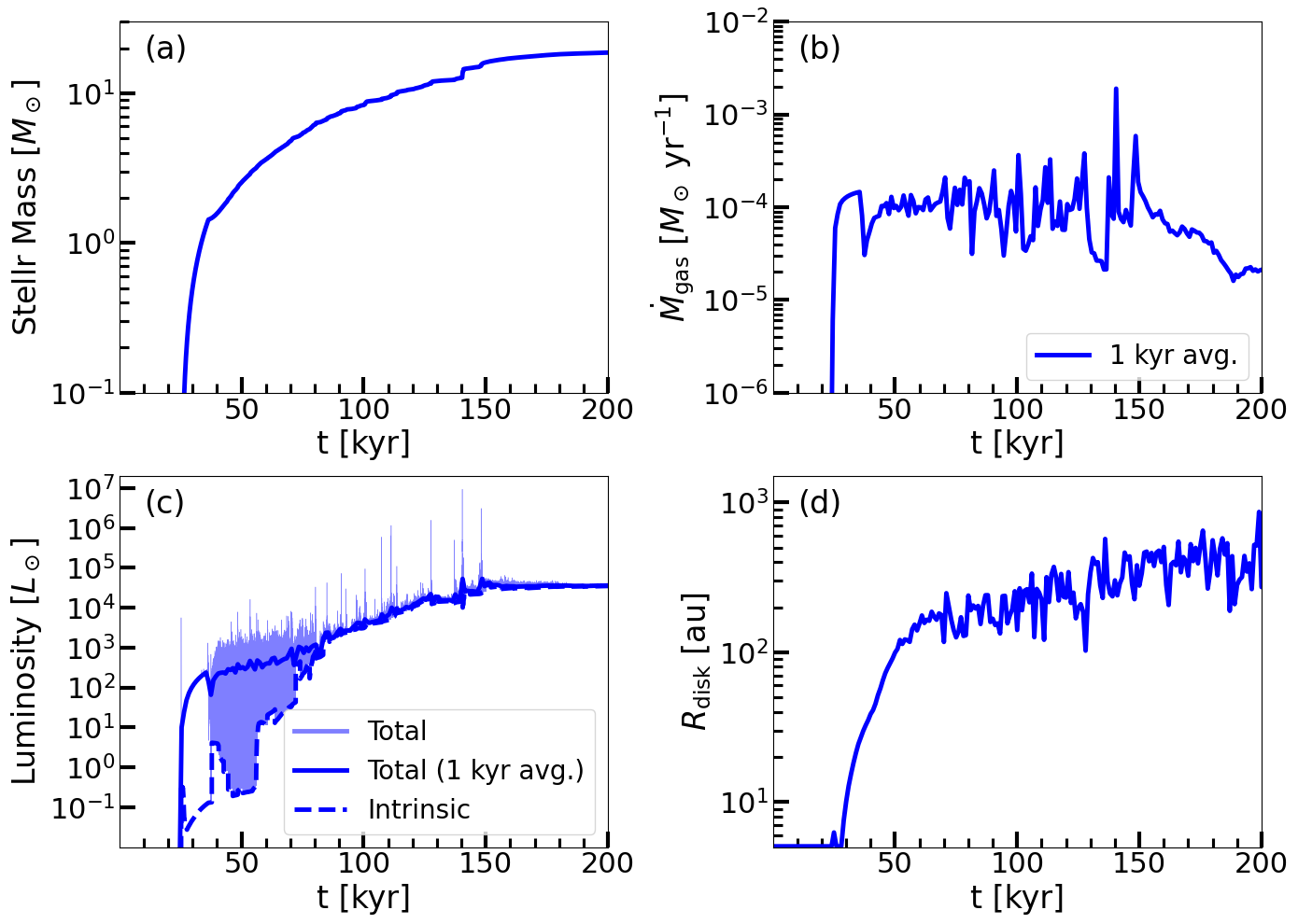}\
\caption{Time evolution of the physical properties of the central star and disk:
(a) stellar mass,
(b) the gas accretion rate to the central star averaged over $1\:\mathrm{kyr}$,
(c) the total and intrinsic stellar luminosities and the total luminosity averaged over $1\:\mathrm{kyr}$, and
(d) the  mean disk radius.
}
\label{Star_Fig_Dust}
\end{figure}

We describe the time evolution of the system using Figure \ref{Star_Fig_Dust}, which illustrates the temporal evolution of the central star and the disk. 
As described in Section \ref{subsec:InitialCondition}, the free-fall time estimated from the initial core conditions is approximately $25\:\mathrm{kyr}$. In the simulation, the central star forms at a comparable epoch of $t = 26\:\mathrm{kyr}$. After its formation, the central star continuously gains mass, as shown by the solid line in panel (a), until $t = 150\:\mathrm{kyr}$, reaching $16\:M_\odot$ and thus exceeding half of the total system mass. This indicates that most of the envelope material has already accreted and that the gas supply to the star begins to decline. As a result, the rate of stellar mass growth gradually slows after this epoch.

This trend is also reflected in the time evolution of the mass accretion rate. To better illustrate the long-term evolution of stellar mass growth, panel (b) shows the accretion rate onto the central star averaged over 1 kyr. Between $t = 26\:\mathrm{kyr}$ and $t = 150\:\mathrm{kyr}$, the accretion rate fluctuates around $\dot{M}_\mathrm{gas} \sim 10^{-4}\:M_\odot\:\mathrm{yr^{-1}}$. During this period, the disk is gravitationally unstable and episodic accretion bursts are triggered by inward migration of gravitationally bound clumps \citep{Meyer2017}. This leads to strong variability in the accretion rate, which can reach values as high as $\sim 3\times10^{-3}\:M_\odot\:\mathrm{yr^{-1}}$. After $t = 150\:\mathrm{kyr}$, the accretion rate gradually declines.

Panel (c) shows the evolution of the stellar luminosity. The total stellar luminosity (light solid line) consists of two components: the intrinsic stellar luminosity (dashed line) and the accretion luminosity generated by mass infall onto the central star. To better illustrate the long-term evolution of stellar luminosity, the total stellar luminosity averaged over 1 kyr is shown as a thick solid line.
At early stages, when the stellar mass is still below the massive protostellar threshold ($< 8\:M_\odot$), the average total luminosity is dominated by accretion luminosity, reaching $\sim10^2$–$10^3\:L_\odot$.
Around $t \sim 80\:\mathrm{kyr}$, when the stellar mass reaches $\sim8\:M_\odot$ and the system enters the massive protostellar phase, the average total luminosity becomes dominated by the intrinsic stellar luminosity and increases to $\sim10^3$–$10^4\:L_\odot$.
After $t \sim 150\:\mathrm{kyr}$, the stellar luminosity continues to increase with stellar mass and eventually approaches a nearly constant value of $\sim4\times10^{4}\:L_\odot$.

The disk is gravitationally unstable until $t \sim 150\:\mathrm{kyr}$, and episodic accretion driven by clump infall produces intense luminosity bursts. The peak luminosity reaches $\sim10^2$–$10^4\:L_\odot$ during the low-mass protostellar phase ($t \lesssim 80\:\mathrm{kyr}$) and increases to $\sim10^5$–$10^7\:L_\odot$ during the massive protostellar phase ($t \lesssim 150\:\mathrm{kyr}$), in agreement with three-dimensional simulations of a similar phenomenon by \citet{Meyer2021}. The details of these burst events in the massive protostellar phase are described in Section \ref{sec:burst}.

To characterize the disk evolution, we describe the temporal evolution of the disk radius. Examples of the identified disk region are shown by the white contours in panels (a.1) and (a.2) of Figure \ref{Gro}. As seen in panel (a.2), the disk becomes strongly nonaxisymmetric due to disk fragmentation, resulting in large azimuthal variations in the disk radius. 
Given the substantial azimuthal fluctuations in the disk radius, a single radius cannot fully characterize the disk structure. We therefore introduce the azimuthally averaged mean disk radius, $R_\mathrm{disk}$, and use its temporal evolution to characterize the disk evolution.

As shown in panel (d), the disk begins to expand almost simultaneously with protostar formation at $t = 26\:\mathrm{kyr}$. Between $t = 26$--$37\:\mathrm{kyr}$, material continues to accrete efficiently onto the central star and the mass accretion rate remains nearly constant. During this stage, the centrifugal radius of the infalling material is smaller than the sink cell radius ($5\:\mathrm{au}$), so the material directly falls into the sink cell and onto the central star. At $t = 37\:\mathrm{kyr}$, the centrifugal radius exceeds $5\:\mathrm{au}$ and a centrifugally supported disk begins to form, marking the onset of the disk formation stage.
Thereafter, the disk continues to expand, reaching $\sim500\:\mathrm{au}$ at $t = 200\:\mathrm{kyr}$. This expansion is driven by continued accretion from the envelope, subsequent viscous spreading, and gravitational interaction of clumps with each other and radial scattering.

\subsection{Gas Radial Distribution}
\label{subsec:gas}

\begin{figure}
\includegraphics[width=\hsize]{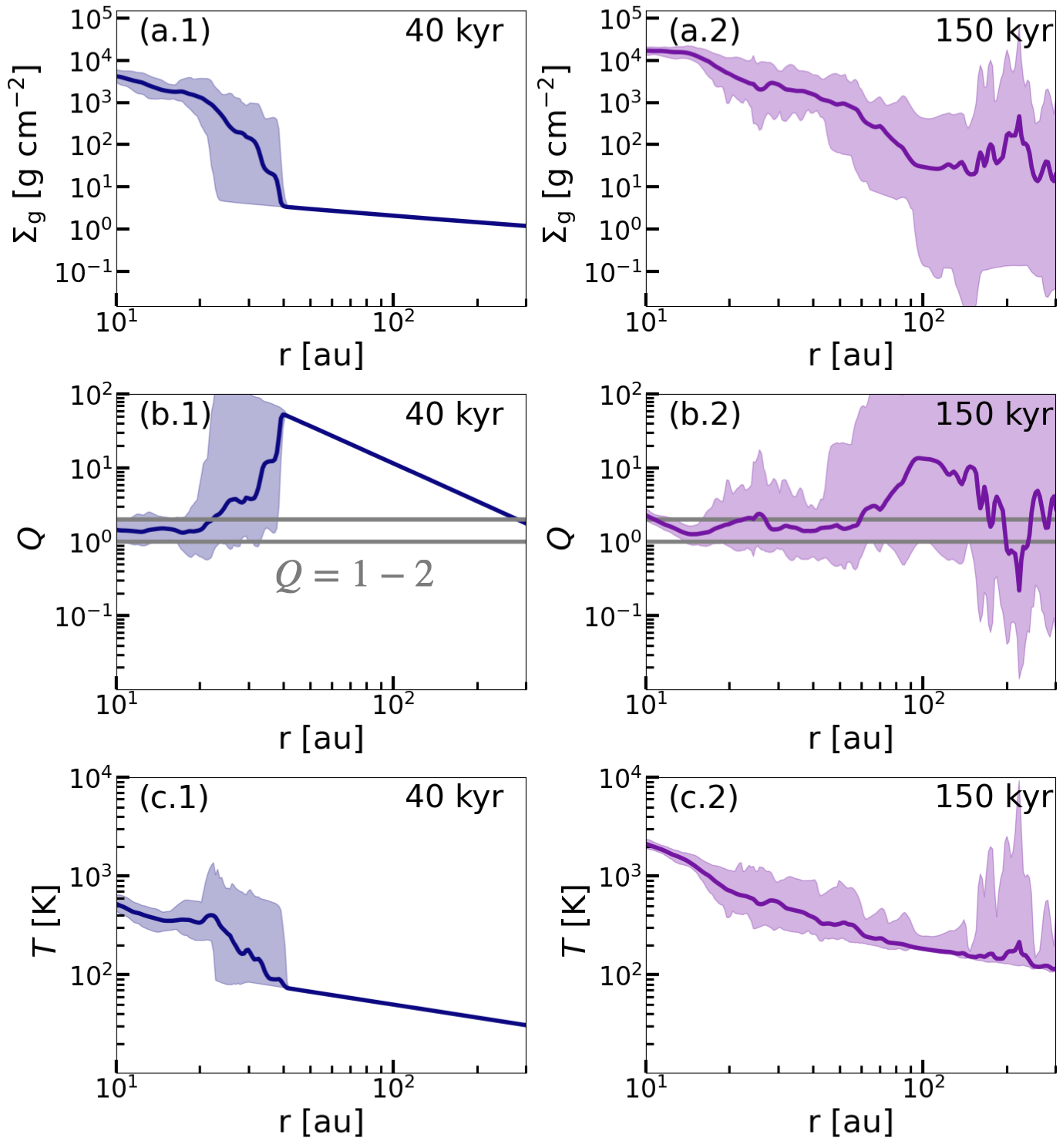}\
\caption{Snapshots of the radial distributions of the gas surface density, the Toomre's $Q$, and the temperature.
The solid lines represent the azimuthally averaged values, where Toomre's $Q$ is mass-weighted, while the shaded areas indicate the maximum and minimum values in the azimuthal direction.
The gray solid lines indicate $Q=1$ and $2$, highlighting the gravitationally unstable regime where spiral structures can develop without necessarily leading to fragmentation.}
\label{1D_Ref}
\end{figure}

Here, we quantitatively describe the gas disk structure using Figure \ref{1D_Ref}. This figure shows the radial profiles of gas physical quantities at two epochs, $t = 40$ and $150\:\mathrm{kyr}$, corresponding to those adopted for the spatial distributions in Figure \ref{Gro}. The solid lines in each panel represent the azimuthally averaged values of the physical quantities, while Toomre's $Q$ is mass-weighted. The shaded regions indicate the range between the maximum and minimum values over azimuth.

As shown by the solid line and the shaded region in panel (a.1), the radial profile of the gas surface density exhibits a high-density, nonaxisymmetric structure within approximately $40\:\mathrm{au}$. The shaded region disappears beyond this radius, indicating that the disk becomes nearly axisymmetric and follows a smooth radial decline. This transition occurs at the disk radius of $\sim40\:\mathrm{au}$ at this epoch (see panel (d) of Figure \ref{Star_Fig_Dust}).
In addition, panel (b.1) shows that the azimuthally averaged Toomre's $Q$ is approximately $1$--$2$ within $\lesssim 20\:\mathrm{au}$, indicating that the disk is gravitationally unstable. 
This instability gives rise to nonaxisymmetric density structures.
At radii of $\sim20$--$40\:\mathrm{au}$, the azimuthally averaged $Q$ exceeds $2$. This elevated value is primarily caused by low-density gap regions between spiral arms that increase the azimuthal average. Nevertheless, the lower bound of the shaded region reaches $Q \lesssim 1$, indicating that the spiral arms are locally gravitationally unstable despite the larger azimuthally averaged value.

The temperature profile in panel (c.1) shows a steep radial gradient around the disk radius. Inside the disk, the high surface density results in a large optical depth ($\tau_\mathrm{R}$), viscous accretion heating and compressional heating raise the azimuthally averaged temperature in the inner disk and producing nonaxisymmetric temperature structures. In contrast, in the envelope, radiative heating from the central star dominates, leading to an axisymmetric temperature distribution with a smooth radial gradient.

We next describe the physical quantities of the disk at $t = 150\:\mathrm{kyr}$. At this stage, the disk has expanded to a radius of roughly $300\:\mathrm{au}$ (panel (b) in Figure \ref{Star_Fig_Dust}). As seen in the gas surface density in panel (a.2), a nonaxisymmetric structure extends across the entire disk.
The distribution of the Toomre's $Q$ in panel (b.2) shows that the mass-weighted azimuthally averaged $Q$ is $\sim1$--$2$ at radii of $\sim10$–$60\:\mathrm{au}$, indicating that this region of the disk is gravitationally unstable, and self-regulated through gravitational torques \citep[e.g,][]{Boley2006}.
At larger radii, $r \gtrsim 60\:\mathrm{au}$, dense spiral arms coexist with low-density inter-arm regions.
In addition, multiple peaks appear in the shaded regions around $\sim200\:\mathrm{au}$, corresponding to clumps formed within the disk.
Strong density contrasts associated with both the spiral structures and clumps produce low-density regions in which the gas surface density decreases drastically.
As a result, Toomre's $Q$ locally reaches values as high as $\sim100$ in these regions.

As the stellar mass increases, in this epoch, the luminosity reaches $\sim 4 \times 10^4\:L_\odot$. As a result, the disk temperature is higher than the state at $t = 40\:\mathrm{kyr}$, as shown in panel (c.2), with temperatures exceeding $\sim 100\:\mathrm{K}$ over most of the disk.
Nonaxisymmetric temperature structures are observed across the entire disk. These features arise from heating associated with viscous accretion and compressional heating. Especially beyond $\sim 100\:\mathrm{au}$, very high temperatures reaching $\sim 10^3$–$10^4\:\mathrm{K}$ are observed. These features result from strong compressional heating within dense clumps.

\subsection{Radial Distribution of the Maximum Grain Radius }\label{sec:Dust}

In this section, we examine how dust growth is regulated in a nonaxisymmetric and time-dependent disk. We show that, in dense disk regions, the maximum grain radius is broadly regulated by the azimuthally averaged fragmentation-limited size, whereas in newly formed outer disk regions and low-density regions, dust growth deviates from local fragmentation equilibrium because of finite collision timescales and dust radial transport timescales.

\begin{figure}
\includegraphics[width=\hsize]{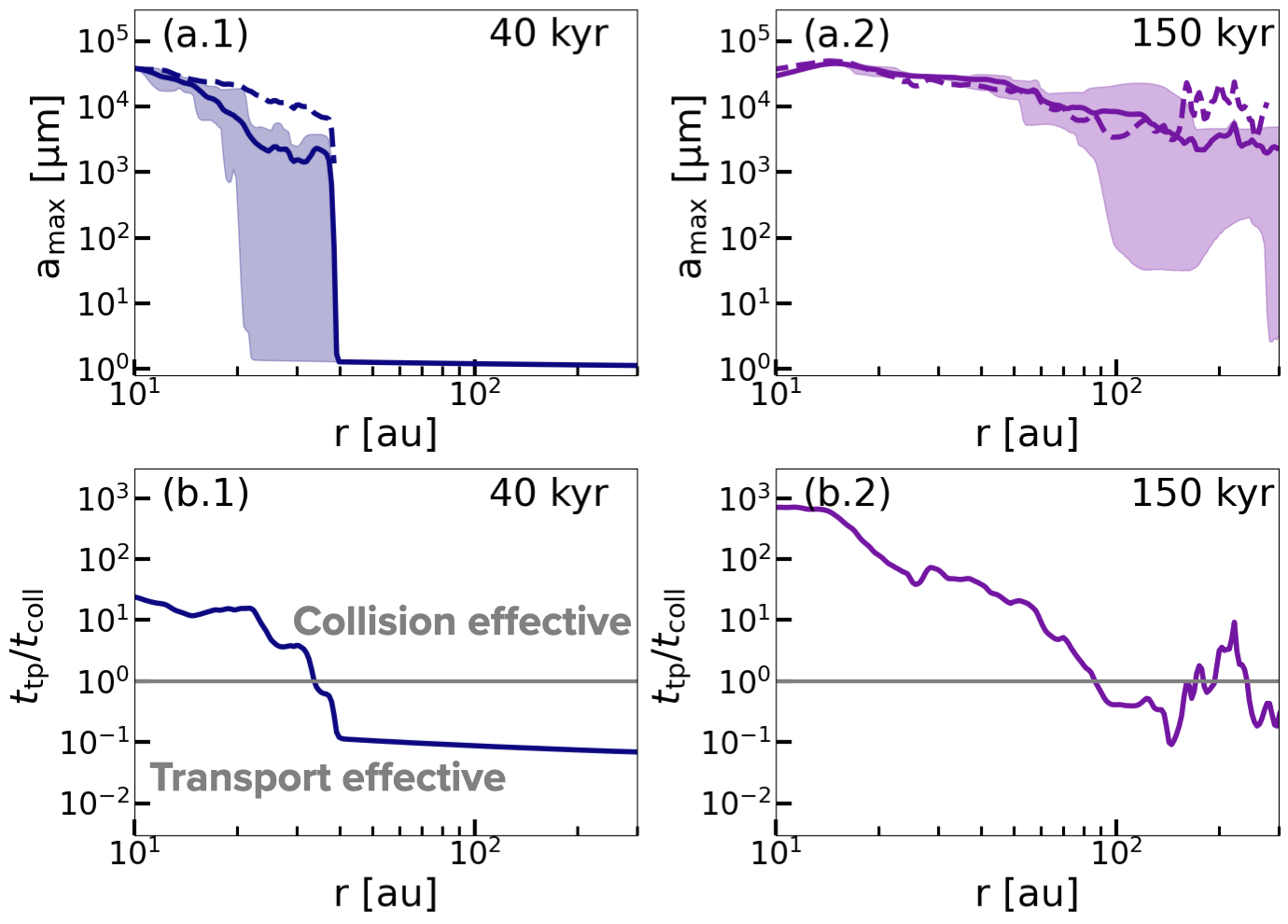}\
\caption{Snapshots of the radial distributions of the maximum grain radius $a_\mathrm{max}$ and the ratio of the transport to collisional timescales $t_\mathrm{tp}/t_\mathrm{coll}$. The solid lines show the azimuthally averaged values, where $a_\mathrm{max}$ is weighted by the dust mass. The shaded regions denote the azimuthal maximum and minimum values. The dashed lines indicate the dust–mass–weighted azimuthal average of the analytical estimate $a_\mathrm{frag,ave}$ from Equation (\ref{Eq:afrag}). The gray solid line indicates $t_\mathrm{tp}/t_\mathrm{coll} = 1$, representing the boundary between collision-effective and transport-effective regimes.}
\label{1D_RefDust}
\end{figure}

Figure \ref{1D_RefDust} shows the radial profiles of the maximum grain radius, $a_\mathrm{max}$, and the ratio of the transport timescale, $t_\mathrm{tp}$, to the collision timescale, $t_\mathrm{coll}$, at $t = 40\:\mathrm{kyr}$ and $150\:\mathrm{kyr}$.
The transport timescale is defined as
\begin{equation}
t_\mathrm{tp} = \frac{r^2 \int_0^{2\pi} \Sigma_\mathrm{g} d\phi}{\dot{M}_\mathrm{gas}},
\end{equation}
where $\dot{M}_\mathrm{gas}$ is the mass transport rate through the disk.
This definition assumes that dust grains remain well coupled to the gas and are transported on the same characteristic timescale as the gas.
We adopt a constant value of $\dot{M}_\mathrm{gas}=1.3\times10^{-4}\:M_\odot\:\mathrm{yr}^{-1}$, corresponding to the typical accretion rate onto the central star during the period from $t=26$ to $150\:\mathrm{kyr}$.
The solid lines in the $a_\mathrm{max}$ profiles represent dust-mass-weighted azimuthal averages, while the shaded regions show the range between the azimuthal minimum and maximum values. Only azimuthally averaged profiles are shown for $t_\mathrm{tp}/t_\mathrm{coll}$.

To characterize the dust growth regime, the dashed lines show the dust-mass-weighted azimuthally averaged fragmentation-limited grain radius, $a_{\mathrm{frag,ave}}$. The fragmentation-limited grain radius is derived following \citet{Binrstiel2012}, based on the drag prescription of \citet{Weidenschilling1977}. 
We define
\begin{equation}
\label{Eq:afrag}
    a_{\mathrm{frag}}=1.8\times\frac{2\Sigma_\mathrm{g} u_\mathrm{frag}^2}{3\pi\rho_\mathrm{s}\alpha c_s^2},
\end{equation}
where the prefactor $1.8$ accounts for the difference in the drag coefficient $C_D$ between our prescription and that of \citet{Weidenschilling1977} \citep[see Figure 5 of][]{Stoyanov2020}.
Since the fragmentation barrier operates only where $u_\mathrm{rel}$ reaches $u_\mathrm{frag}$ and collisions occur sufficiently frequently, $a_{\mathrm{frag,ave}}$ is shown only within the disk radius.

At $t = 40\:\mathrm{kyr}$, immediately after disk formation, panel (a.1) shows dust grains in the disk region ($\sim40\:\mathrm{au}$) have already grown significantly from the initial maximum grain radius of $1\:\mathrm{\mu m}$, reaching up to $\sim10^4\:\mathrm{\mu m}$. 
{Such rapid dust growth is consistent with previous studies of dust evolution in low-mass protostellar systems \citep[e.g.,] []{Lebreuilly2023,Tsukamoto2023,Vorobyov2025}.}
As shown in panel (b.1), $t_\mathrm{coll}$ is shorter than $t_\mathrm{tp}$ in the disk region, indicating that dust grains can grow through collisions faster than dust transport inward.
A comparison of the azimuthally averaged $a_\mathrm{max}$ and $a_\mathrm{frag,ave}$ in panel (a.1) shows that $a_\mathrm{max}$ is nearly identical to $a_\mathrm{frag,ave}$ within $\sim20\:\mathrm{au}$, indicating that dust growth there has already entered the fragmentation-limited regime.
In contrast, at $\sim20$--$40\:\mathrm{au}$, $a_\mathrm{max}$ remains smaller than $a_\mathrm{frag,ave}$.
In this region, reaching $a_\mathrm{frag,ave}$ requires $t_\mathrm{coll} \sim 1\:\mathrm{kyr}$.
However, the disk is still continuously expanding outward, and the $\sim20$--$40\:\mathrm{au}$ region has existed for only $\sim1\:\mathrm{kyr}$ since becoming part of the disk.
This timescale is comparable to the collision timescale required for dust grains to grow to the fragmentation barrier.
As a result, dust grains in this newly formed outer disk region have not yet had sufficient time to grow up to $a_\mathrm{frag,ave}$.

The shaded region of $a_\mathrm{max}$ represents azimuthal variations caused by nonaxisymmetric structures within the disk. This is because the disk develops spiral arms, where the enhanced dust surface density increases the collision frequency and promotes efficient local dust growth. By contrast, in the surrounding low-density regions, dust growth remains inefficient, and grain sizes increase only marginally from the initial $1\:\mathrm{\mu m}$ to at most a few times larger.

In the envelope, dust growth does not occur. As shown in panel (b.1), $t_\mathrm{tp}/t_\mathrm{coll} < 1$ in this region, reflecting both the low dust surface density that leads to long collision timescales and the rapid infall of dust into the inner disk before substantial growth can occur. In low-mass clouds, on the contrary, mild dust growth takes place already in the pre-stellar stage \citep{Vorobyov2025}.

As time progresses to $t = 150\:\mathrm{kyr}$, panel (a.2) shows that dust growth extends across the entire disk as the disk expands. Within the inner region $\lesssim60\:\mathrm{au}$, $a_\mathrm{max}$ exhibits an approximately axisymmetric distribution, even though the gas surface density (panel (a.2) in Figure \ref{1D_Ref}) still shows pronounced nonaxisymmetric structures.
At this stage, approximately $\sim100\:\mathrm{kyr}$ have elapsed since disk formation, which is significantly longer than the collisional timescale ($t_\mathrm{coll} \sim \mathrm{kyr}$).
In addition, $t_\mathrm{tp}/t_\mathrm{coll} \gg1$, and $t_\mathrm{coll}$ exceeds the local Keplerian timescale. Since the gas properties vary significantly between spiral-arm and inter-arm regions and evolve on orbital timescales, dust grains experience a wide range of local conditions before undergoing collisions. Consequently, dust growth reflects the cumulative effect of these variations rather than the instantaneous local environment. This leads to a maximum grain radius that is more closely related to the azimuthally averaged fragmentation condition, i.e., $a_\mathrm{max}\sim a_\mathrm{frag,ave}$.

In the outer disk ($\sim60$--$200\:\mathrm{au}$), the azimuthally averaged $a_\mathrm{max}$ generally agrees with $a_{\mathrm{frag,ave}}$. In this region, the disk develops more prominent nonaxisymmetric structures than in the inner regions, including spiral arms, dense clumps and the gap structures. As a result, high-density regions that contain most of the mass, such as spiral arms and clumps, dominate the mass-weighted azimuthal averages of $a_\mathrm{max}$ and $a_{\mathrm{frag,ave}}$. Within these dense regions, frequent dust collisions lead to $a_\mathrm{max}$ being regulated by collisional fragmentation, yielding values close to $a_{\mathrm{frag,ave}}$.

However, the shaded region of $a_\mathrm{max}$ clearly indicates significant nonaxisymmetry in this part of the disk. This asymmetry originates from the strong gas surface density contrast, which exceeds two orders of magnitude between spiral arms or clumps and the surrounding low-density regions.
In the high-density spiral arms and clumps, dust grains efficiently grow through collisional coagulation, producing large grains. By contrast, the low-density regions contain not only grains formed through local collisional growth, but also grains leaked from the dense spiral arms and clumps into the surrounding low-density regions. 
As spiral arms and clumps continuously form and dissipate in the gravitationally unstable disk, fluid elements containing previously grown dust grains can be transported from these dense structures into the surrounding low-density regions.
Some of these leaked grains exceed the local fragmentation barrier in the low-density regions.
Although such grains are larger than the local fragmentation-limited size, the collision frequency in the low-density regions is too low for fragmentation to efficiently reduce their sizes.
As a result, these large grains can survive for extended periods, causing $a_\mathrm{max}$ to significantly deviate from the value expected from $a_\mathrm{frag,ave}$.

Finally, at $r \sim 200$--$300\:\mathrm{au}$, the azimuthally averaged $a_\mathrm{max}$ is smaller than $a_{\mathrm{frag,ave}}$. This is due to $t_\mathrm{tp}/t_\mathrm{coll} \lesssim 1$, where infrequent collisions prevent dust grains from growing to the fragmentation limit.

\section{Effects of Dust Growth on Disk Evolution} \label{sec:CompareDG}

In this section, we investigate the effects of dust growth on the evolution of massive protostellar disks and their central stars. Dust growth regulates the Rosseland and Planck mean opacities. As dust grains grow, the Rosseland mean opacity $\kappa_\mathrm{R}$ and the Planck mean opacity $\kappa_P$ decrease at $T\gtrsim100\:\mathrm{K}$.
However, since the disk remains optically thick over most of its extent throughout the main accretion phase, radiative cooling is primarily controlled by the Rosseland mean opacity, as can be seen from Equation (\ref{eq:cooling}). Therefore, the reduction in $\kappa_\mathrm{R}$ enhances radiative cooling, leading to lower disk temperatures. This, in turn, strengthens the gravitational instability of the disk.

To demonstrate this effect, we compare a massive protostellar disk model that includes dust growth (hereafter the Dust Growth model; DG) with a model that does not consider dust growth (hereafter the No Dust Growth model;  NDG).
The DG model is identical to the model described in Section \ref{sec:Result}.
In the NDG model, the maximum grain radius is fixed at $a_\mathrm{max}=1\:\mathrm{\mu m}$ at all times, resulting in a higher Rosseland mean opacity $\kappa_\mathrm{R}$ than in the DG model (Figure \ref{DustOpa}). Consequently, the disk temperature is systematically higher. All other physical conditions are identical between the two models.
By comparing these two models, we highlight the differences in disk evolution and the associated consequences of dust growth.

\subsection{Comparison of Models with and without Dust Growth}
\label{subsec:no-dustgrowth}

\begin{figure*}
\centering
\includegraphics[width=0.8\textwidth]{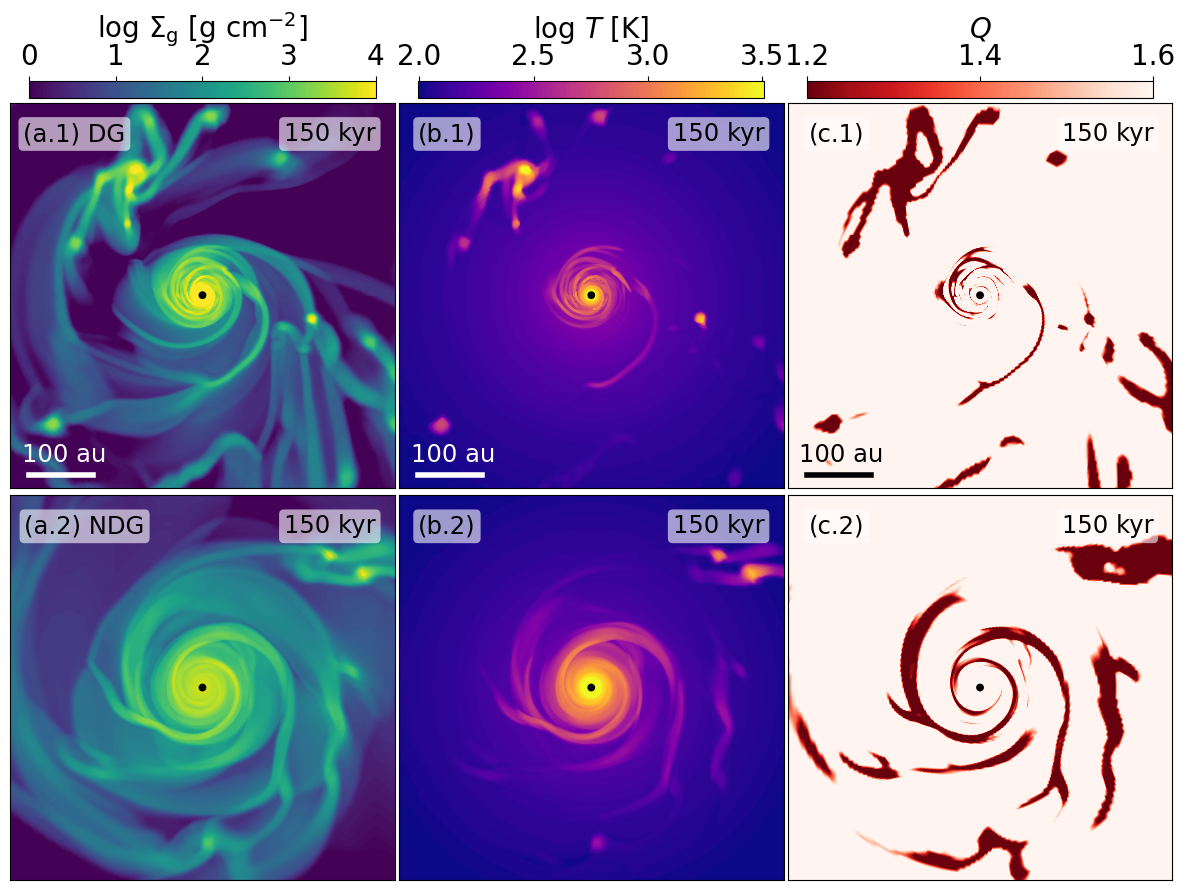}
\caption{Snapshots of the gas surface density, temperature, and Toomre's $Q$ within the inner $600 \times 600\:\mathrm{au}^2$ region for the Dust Growth (upper panels) and No Dust Growth (lower panels) models. }
\label{CompareNoGro}
\end{figure*}

We present a comparative analysis of the DG and NDG models. Figure \ref{CompareNoGro} shows the spatial distributions of key physical quantities at $t = 150\:\mathrm{kyr}$ for both models. From left to right, the panels display the gas surface density, temperature, and Toomre's $Q$.

The gas surface density distributions in panels (a.1) and (a.2) show that the DG model undergoes multiple self-gravitational fragmentation events, whereas the NDG model develops a more extended disk region of $\sim 300\:\mathrm{au}$ with less fragmentation.

In the DG model, the opacity is one to two orders of magnitude lower than in the NDG model, enabling more efficient radiative cooling of the heat generated by viscous accretion heating and compressional heating. As a result, the temperature within $\lesssim200\:\mathrm{au}$ is systematically lower in the DG model than in the NDG model, as shown in panels (b.1) and (b.2). This enhanced cooling strengthens gravitational instability, thereby promoting fragmentation of spiral arms \citep{Gammie2001,Rice2003,Baehr2015}.
In contrast, cooling is less efficient in the NDG model, making spiral arms less prone to fragmentation. Consequently, the spiral-arm regions with Toomre's $0.6\lesssim Q \lesssim 1.4$ extend to $\sim200\:\mathrm{au}$ (panel c.2), significantly larger than the $\sim60\:\mathrm{au}$ region seen in the DG model (panel c.1).

\begin{figure}
\centering
\includegraphics[width=0.5\textwidth]{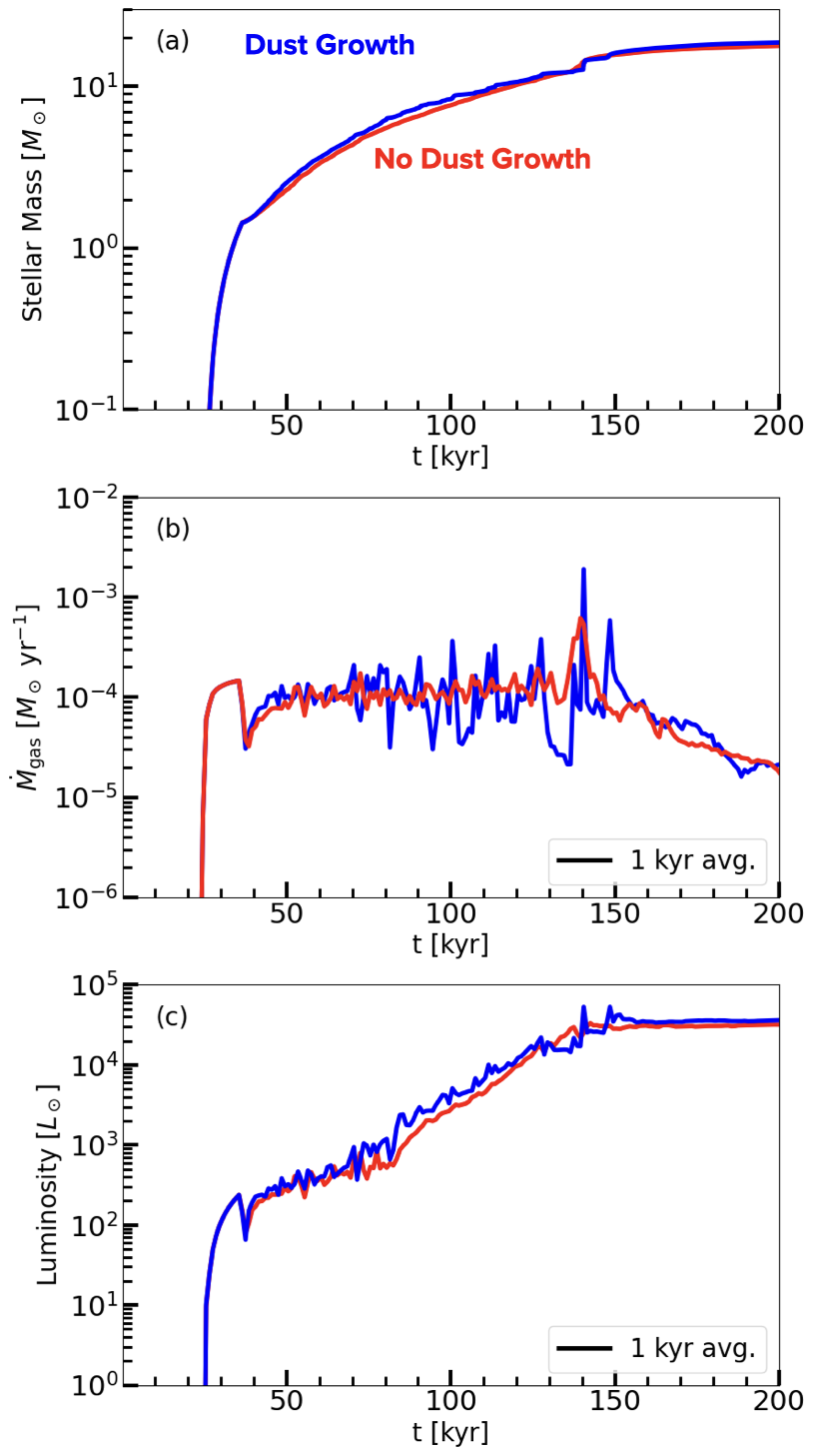}\
\caption{Time evolution of the physical properties of central star and disk for the Dust Growth model (blue lines) and the No Dust Growth model(red lines):
(a) stellar mass,
(b) the gas accretion rate to the central star averaged over $1\:\mathrm{kyr}$,
(c) the total stellar luminosity averaged over $1\:\mathrm{kyr}$.
}
\label{Star_Fig}
\end{figure}

To compare the evolution of the two systems, Figure \ref{Star_Fig} presents the time evolution of the central star mass, accretion rate on to the central star, and the total stellar luminosity.
Since the initial conditions are identical except for the treatment of dust growth, the stellar growth histories and final stellar masses (panel a) are nearly identical between the two models. Correspondingly, the final stellar masses are also similar, with $19\:M_\odot$ for the DG model and $18\:M_\odot$ for the NDG model.

The mass accretion rate (panel b) and stellar luminosity (panel c) exhibit broadly similar temporal evolution in both models; however, the DG model shows significantly larger variability.
These differences arise from more efficient cooling in the DG model, which makes the disk more gravitationally unstable, thereby promoting more frequent disk fragmentation.
As a result, episodic accretion events driven by clump infall occur more frequently and with larger amplitudes.

\subsection{Accretion Variability and Bursts by Dust Growth}
\label{sec:burst}
In this section, we examine how dust growth affects the temporal variability of mass accretion and luminosity outbursts.
Although the global evolution of the mass accretion rate onto the central star is broadly similar in the DG and NDG models, dust growth enhances gravitational instability and induces stronger episodic variability in the accretion rate. To investigate this behavior in more detail, we analyze the accretion rate on yearly timescales and compare the two models.

\begin{figure*}
\centering
\includegraphics[width=0.9\textwidth]{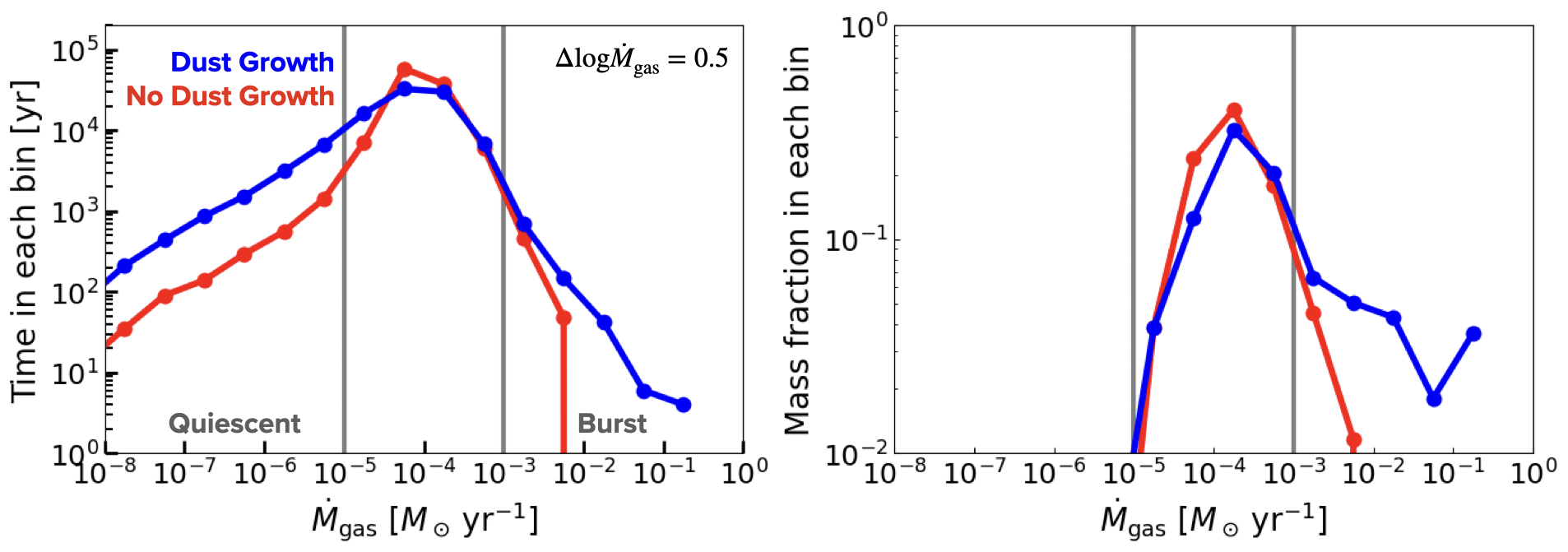}\
\caption{Detailed analysis of the gas accretion rate onto the central star, $\dot{M}_\mathrm{gas}$, for both the Dust Growth model (blue) and the No Dust Growth model (red) during the main accretion phase.
(left) Cumulative time in each $\dot{M}_\mathrm{gas}$ bin, where the bin width is set to $\Delta \log \dot{M}_\mathrm{gas} = 0.5$.
(right) Cumulative mass fraction to central stellar mass derived in each $\dot{M}_\mathrm{gas}$ bin.
In this study, we define the quiescent phase as $\dot{M}_\mathrm{gas} < 10^{-5}\: M_\odot\:\mathrm{yr^{-1}}$ and the burst phase as $\dot{M}_\mathrm{gas} > 10^{-3}\: M_\odot\:\mathrm{yr^{-1}}$.}
\label{Acumula}
\end{figure*}

Panel (a) of Figure \ref{Acumula} shows the cumulative time spent in different bins of the gas mass accretion rate. The accretion rate, $\dot{M}_\mathrm{gas}$, is divided into bins of width $\Delta \log \dot{M}_\mathrm{gas} = 0.5$. The analysis focuses on the main accretion phase, from $t = 37$ to $150\:\mathrm{kyr}$, during which the central star accretes most of its mass (see Section \ref{subsec:Global} and \ref{subsec:no-dustgrowth}). In both models, the accretion rate spends most of the time around $\sim 10^{-4}\:M_\odot\:\mathrm{yr}^{-1}$, consistent with the typical values shown in panel (b) of Figure \ref{Star_Fig}.

We define accretion rates exceeding this typical value by about an order of magnitude as the burst phase. Under this phase, the maximum accretion rate differs significantly between the two models. In the DG model, the peak accretion rate reaches $0.2\:M_\odot\:\mathrm{yr}^{-1}$, whereas it is limited to $6\times10^{-3}\:M_\odot\:\mathrm{yr}^{-1}$ in the NDG model. 
Moreover, the total duration of such high-accretion states is longer in the DG model, amounting to $880\:\mathrm{yr}$ compared to $510\:\mathrm{yr}$ in the NDG model.

Strong bursts are accompanied by periods of very low accretion. In this study, phases with $\dot{M}_\mathrm{gas} \leq 10^{-5}~M_\odot~\mathrm{yr}^{-1}$ are defined as quiescent phases. The cumulative duration of the quiescent phase is $27~\mathrm{kyr}$ in the DG model, indicating that approximately $25\:\%$ of the main accretion phase is spent in a low-accretion state when dust growth is included. In contrast, the corresponding duration is only $4~\mathrm{kyr}$ in the NDG model.

To quantify how stellar mass is accumulated at different accretion rates, panel (b) of Figure \ref{Acumula} shows the fraction of the final stellar mass accreted at each accretion rate. In both models, the largest fraction of stellar mass is accreted at $\dot{M}_\mathrm{gas} \sim 10^{-4}\:M_\odot\:\mathrm{yr}^{-1}$. In contrast, the burst phase contributes about $20\:\%$ of the total stellar mass in the DG model, whereas its contribution is only $\sim6\:\%$ in the NDG model.

\begin{figure}
\centering
\includegraphics[width=0.4\textwidth]{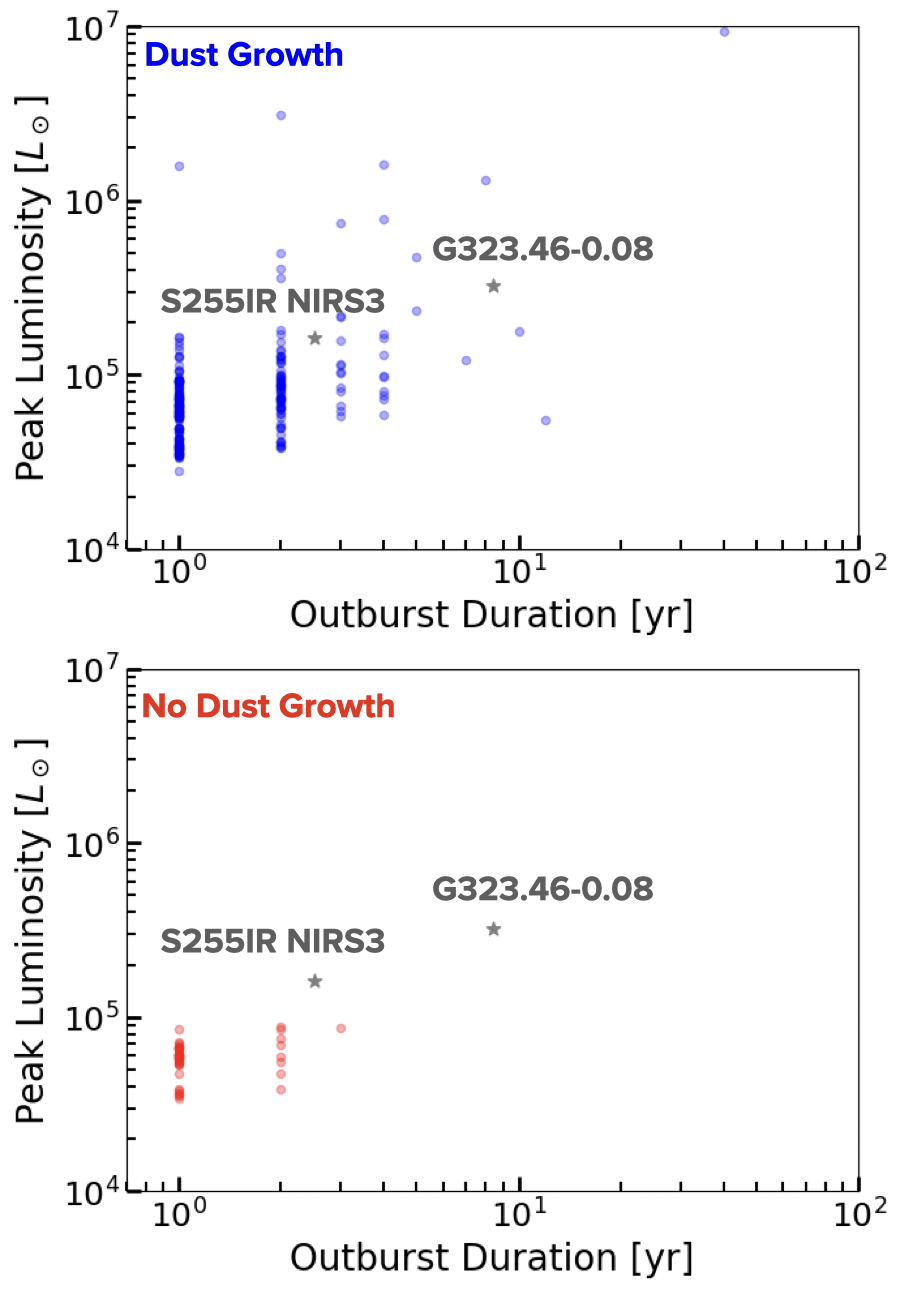}\
\caption{Peak luminosity and outburst duration during the main accretion phase for the Dust Growth model (upper) and the No Dust Growth model (lower).
An outburst is defined as a period during which the total luminosity exceeds 2.5 times the intrinsic stellar luminosity.
For comparison, the observed peak luminosities and outburst durations of S255IR NIRS3 and G323.46-0.08 are shown as gray markers \citep{Szymczak2018,Wolf2024A&A}.
}
\label{Burst}
\end{figure}

Finally, we examine the properties of luminosity bursts. Figure \ref{Burst} shows their peak luminosities and durations. A luminosity burst is defined as a period during which the stellar luminosity exceeds $10^4\:L_\odot$ and is more than 2.5 times higher than the intrinsic stellar luminosity. In the DG model, 248 bursts occur with peak luminosities of $\sim10^4$--$10^7\:L_\odot$ and durations ranging from one year to several tens of years. In contrast, the NDG model produces 44 bursts with peak luminosities of $\sim10^4$--$10^5\:L_\odot$ and durations of 1–3 years. These results indicate that enhanced gravitational instability due to dust growth increases the frequency, duration, and intensity of accretion-driven luminosity bursts.

\section{Discussion}\label{sec:Dis}
In the preceding sections, we presented the coevolution of dust growth and massive protostellar disks.
Here, we first discuss how dust growth affects the evolution of massive protostellar disks, with particular emphasis on disk fragmentation and episodic accretion.
We then investigate how the evolving disk structure, in turn, influences dust growth.

\subsection{Disk Fragmentation and Episodic Accretion Enhanced by Dust Growth}

Our simulations demonstrate that dust growth can significantly influence the evolution of massive protostellar disks.
Dust grains rapidly grow from interstellar sizes, reaching centimeter scales within $\sim\mathrm{kyr}$ during the early stages of disk formation, implying that dust evolution affects disk dynamics from the earliest phases. dust growth reduces the infrared opacity and enhances radiative cooling, which strengthens gravitational instability in the disk. As a consequence, disks with dust growth exhibit more frequent fragmentation and episodic accretion events than models without dust growth. These processes also produce observable signatures in the form of accretion-driven luminosity outbursts.

As shown in Figure \ref{Burst}, the model with dust growth produces more frequent, longer-lasting, and more luminous bursts than the model without dust growth. It is therefore useful to compare these burst properties with observed outbursts in massive protostars. A variety of outburst events from massive protostars have been reported to date \citep[e.g.,][]{Caratti2017,Fedriani2023,Xu2025,Kumar2026,Liu2026}. Among them, luminosity bursts observed in S255IR NIRS3 and G323.46$-$0.08 last for several years and reach peak luminosities exceeding $10^5\:L_\odot$ \citep{Szymczak2018,Wolf2024A&A}. These durations and luminosities are broadly consistent with those obtained in the dust-growth model. This agreement suggests that dust-growth-enhanced gravitational instability may contribute to producing luminosity bursts with properties comparable to those observed in massive protostars.

Because disk fragmentation is a plausible pathway to the formation of close multiple systems, our results may also have implications for the origin of massive binary and multiple systems.
Most massive stars form as members of binary or higher-order multiple systems \citep{Offner2023}.
In particular, systems with separations of $\lesssim1000\:\mathrm{au}$ are commonly thought to originate from disk fragmentation, rather than fragmentation of the parent molecular cloud \citep{Kuruwita2023,Pandian2024,Qiuyi2026}. Numerous theoretical studies have investigated disk fragmentation processes \citep[e.g.,][]{Kratter2010,Meyer2018,Oliva2020A,Risse2023}, and observational evidence supporting multiple star formation through disk fragmentation has also emerged in recent years \citep[e.g.,][]{Beltran2016,Ilee2018,Izquierdo2018,Zhang+2019,Tanaka2020,Ahmadi2023,Shanghuo2025Nat}. Our results suggest that dust growth may further promote the formation of massive multiple systems by promoting disk fragmentation.

It should be emphasized, however, that our results do not imply that dust growth alone is responsible for triggering outbursts. Previous studies have shown that similar bursts can arise even without dust growth, depending on the initial conditions of the parent molecular cloud core, including high core masses that produce large accretion rates or sufficiently large angular momentum \citep{Meyer2021}.
In addition, several processes not included or not fully resolved in our simulations may affect outburst activity, including boundary-layer or magnetospheric accretion, thermal instability, and tidal disruption of clumps in the inner disk \citep{Kley1996,Romanova2009,Bell1994,Elbakyan2024,Elbakyan2025,Elbakyan2026}.
Despite these caveats, our results suggest that dust growth facilitates the development of gravitational instability and thereby promotes episodic accretion activity.

We note that the strength of the dust-growth feedback depends on the adopted fragmentation threshold velocity, $u_\mathrm{frag}$. In this study, we adopt $u_\mathrm{frag}=10\:\mathrm{m \:s^{-1}}$, following previous studies, which allows dust grains to grow to centimeter scales. However, the fragmentation threshold velocity remains highly uncertain. For smaller values of $u_\mathrm{frag}$, dust growth becomes less efficient. Nevertheless, even in models with $u_\mathrm{frag}=1\:\mathrm{m\:s^{-1}}$, where the maximum grain radius reaches only $a_\mathrm{max}\sim100\:\mathrm{\mu m}$, the dust opacity still decreases sufficiently to enhance gravitational instability. As a result, disk fragmentation and luminosity outbursts comparable to those inferred from observations can still occur.

\subsection{Dust Growth in Massive Protostellar Disks}
\label{Compare1D}

We discuss how dust growth is regulated in dynamically evolving, nonaxisymmetric massive protostellar disks. A key result of this study is that the maximum grain radius in massive protostellar disks can often be estimated using simple steady, axisymmetric dust-growth models \citep{Yamamuro2023,Yamamuro2025}, even when the disks are strongly nonaxisymmetric and time dependent.

As described in Section \ref{sec:Dust}, the maximum grain radius, $a_\mathrm{max}$, becomes approximately axisymmetric and approaches the azimuthally averaged fragmentation-limited size, $a_\mathrm{frag,ave}$. This behavior is evident at radii of $r\sim10$--$60\:\mathrm{au}$, where $t_\mathrm{tp}/t_\mathrm{coll}\gtrsim10$ (panel (b.2) of Figure~\ref{1D_RefDust} ). In contrast, in the outer disk where $t_\mathrm{tp}/t_\mathrm{coll}\lesssim1$, the local distribution of $a_\mathrm{max}$ becomes strongly nonaxisymmetric. In these outer regions, dust grains are transported inward before experiencing enough collisions to approach the fragmentation-limited size. The transition occurs only when $t_\mathrm{tp}/t_\mathrm{coll}$ significantly exceeds unity, because multiple collisions are required for grains to grow substantially and reach $a_\mathrm{frag,ave}$.

These results suggest that dust growth is governed primarily by the ratio $t_\mathrm{tp}/t_\mathrm{coll}$ rather than by the global morphology of the disk. When this ratio is sufficiently large, dust grains undergo enough collisions to approach the local fragmentation-limited size before being transported by the evolving disk flow. As a result, the grain-size distribution becomes relatively insensitive to transient nonaxisymmetric structures and is instead regulated by the local balance between dust growth and fragmentation. This explains why steady, axisymmetric dust-growth models can provide reasonable estimates of the maximum grain radius even in dynamically evolving massive protostellar disks.

Recent observations have revealed pronounced nonaxisymmetric structures in massive protostellar disks, including spiral arms and fragmented structures \citep[e.g.,][]{Beuther2017,Motogi2019,Johnston2020,Sanhueza2021,Burns2023}. Our results suggest that simple steady, axisymmetric dust-growth models can still provide useful estimates of grain sizes in dense regions of such disks, even when the global disk structure is dynamically evolving and nonaxisymmetric.
Therefore, the applicability of steady, axisymmetric dust-growth models depends not on the global morphology of the system, but on whether dust grains can undergo sufficient collisions before being transported away.

Interpreting observed spiral structures nevertheless requires caution. Spiral-like features in massive protostellar systems are not always generated by disk self-gravity and may instead trace infalling streamers \citep{Fernand2023,Olguin2025,Weng2026}. In contrast to self-gravitating spiral arms, dust grains within streamers may not experience sufficiently frequent collisions for efficient dust growth because of their rapid infall and relatively low densities. Consequently, grain-growth properties inferred for self-gravitating disks may not be directly applicable to streamer-dominated structures.

\subsection{Future Works and Caveats}

{{In this study, we investigate the coevolution of dust growth and massive protostellar disks using a two-dimensional thin-disk framework. Recently, a fully three-dimensional version of the FEOSAD code, ngFEOSAD, has been developed \citep{2024A&A...683A.202V}. Although radiative thermal evolution as considered in this work is not yet implemented in ngFEOSAD, future developments of the code will provide an opportunity to test the predictions of our two-dimensional simulations with fully three-dimensional simulations.}}

{{A more comprehensive understanding of coevolution of dust growth and massive protostellar disks requires the inclusion of feedback processes such as radiation pressure, photoevaporation, and magnetically driven disk outflow \citep[e.g.,][]{Tanaka+2017Apj,Rosen2022}. By incorporating these feedback processes, we can investigate how effectively the disk can shield accreting material from such feedback and, consequently, how much material can accrete from the disk onto the protostar.
Furthermore, the change in dust opacity associated with dust growth is expected to affect the efficiency of feedback from radiation pressure and photoevaporation \citep[e.g.,][]{Kuiper2018,Mignon2020}.
Dust growth may also alter the ionization degree of the disk through the adsorption of charged particles onto dust grains, thereby affecting the evolution of the disk magnetic field through nonideal MHD effects, as reported in studies of low-mass protostellar disks \citep[e.g.,][]{Lebreuilly2023,Tsukamoto2023}.
Such changes in the magnetic field may also influence the launching and evolution of magnetically driven disk outflows in massive protostellar disks \citep[e.g.,][]{Commercon2022,Oliva2023}.
Future studies incorporating these processes, together with radiation pressure and photoevaporation, will be important for exploring the broader coevolution of dust growth and massive protostellar disks.}}

{We acknowledge that there are still uncertainties associated with the evolution of dust composition, grain shape, porosity, and other dust properties \citep{Semenov2003,Carphine2025}. However, as shown in Figure \ref{DustOpa}, dust growth can change the dust opacity by orders of magnitude, whereas variations associated with dust composition and grain shape are expected to be limited to factors of a few. Thus, although these uncertainties may affect the quantitative value of the opacity, the opacity change associated with dust growth is expected to remain the dominant effect.}

{{In this study, we assume a fixed power-law index of $q=3.5$ for the dust size distribution. This assumption may break down due to the coagulation of small grains with radius of $\lesssim 1\:\mathrm{\mu m}$ driven by Brownian motion \citep[e.g.,][]{Dullemond2005}. However, these very small grains are substantially smaller than the wavelengths relevant to the thermal emission considered here, $\lambda_\mathrm{peak}\sim10\:\mathrm{\mu m}$, and their contribution to the overall variation in opacity is therefore expected to be limited (see Figure 2 of \cite{Vorobyov2026}). In contrast, the dust growth considered in this study increases $a_\mathrm{max}$ from $\sim 1\:\mathrm{\mu m}$ to several orders of magnitude larger sizes, resulting in a substantial change in the dust opacity. Therefore, the effect of coagulation among the smaller grains is expected to be secondary compared with the opacity change associated with the dust growth in this work.}}

{Recently, bouncing has been proposed as a possible outcome of collisions between fluffy dust aggregates. However, the outcome of such collisions depends on several factors, including the fluffiness of the dust aggregates, collision velocity, and grain radius, making it a highly complex process \citep[e.g.,][]{Michoulier2024,Oshiro2025,Arakawa2026}. Furthermore, when fluffy dust aggregates undergo repeated collisions, the evolution of their fluffiness can further complicate the dust-growth process. Thus, although bouncing may be a physically important process, its treatment is associated with substantial uncertainties. For this reason, we do not include bouncing in this study. If bouncing is effective, the amount of small dust fragments produced through collisions could be reduced \citep{Domink2024A}. This could decrease the abundance of small grains that are sensitive to infrared wavelengths, leading to a further reduction in opacity.
This would support our conclusion that dust growth enhances the gravitational instability of massive protostellar disks.}

\section{Summary} \label{sec:sum}

In this study, we have investigated dust growth and its impact on the evolution of massive protostellar disks. Dust growth affects the observational interpretation of massive protostellar disks, and also modifies the cooling efficiency of the disk. To explore the coevolution of dust growth and disk evolution, we performed hydrodynamical simulations using the FEOSAD code. Our simulations self-consistently follow the gravitational collapse of an initially rotating molecular cloud core, leading to the formation and evolution of a central star and its circumstellar disk, while simultaneously tracking dust growth within the disk. In addition, we incorporated the evolution of the dust opacity associated with dust growth, thereby accounting for changes in the cooling efficiency.

We found that in massive protostellar disks, dust grains rapidly grow from interstellar sizes of $\sim1\:\mathrm{\mu m}$ to centimeter scales within a timescale of $\sim\mathrm{kyr}$ shortly after disk formation. 
This rapid growth implies that dust evolution continuously influences disk evolution from the earliest stages. As dust grains grow, the dust opacity decreases, enhancing radiative cooling and lowering the disk temperature, which can promote the development of gravitational instability.

We compared the evolution of the protostar-disk system with and without dust growth. Enhanced cooling due to dust growth strengthens gravitational instability in the disk, leading to disk fragmentation and the onset of episodic accretion. As a result, mass accretion onto the central star is intensified, increasing the frequency, duration, and magnitude of luminosity outbursts. The outbursts obtained in our dust growth simulation reached peak luminosities up to $\sim10^7\:L_\odot$. At the same time, the simulation also produced bursts with luminosities of $\sim10^5\:L_\odot$ and durations of several years, comparable to observationally reported quantities. Furthermore, disk fragmentation induced by dust growth may contribute to the formation of binary or multiple systems, which are commonly observed around massive stars.

Finally, we have examined the detailed behavior of dust growth in massive protostellar disks, which exhibit nonaxisymmetric and time-variable structures such as spiral arms and clumps driven by gravitational instability. We find that, in sufficiently dense regions where the collisional timescale is shorter than the transport timescale, the maximum grain radius is primarily regulated by azimuthally averaged collisional fragmentation and therefore remains approximately axisymmetric despite the nonaxisymmetric gas structures. This suggests that the fragmentation-limited dust growth framework developed for steady, axisymmetric disks remains broadly applicable even in dynamically evolving self-gravitating disks.

\section*{}
The authors are grateful to Satoshi Okuzumi for useful discussions.
The authors are thankful to the anonymous referee for insightful comments that helped improve the manuscript.
This work was supported by JSPS KAKENHI Grant Numbers 24KJ1092 and 25K07365, and by the NAOJ ALMA Scientific Research Grant, Code 2025-29B.
Numerical computations were carried out on Cray XC50 and XD2000 at Center for Computational Astrophysics, National Astronomical Observatory of Japan.
We acknowledge support from the AAS Publication Support Fund, which covered the publication charges for this work.

\bibliography{ms}{}
\bibliographystyle{aasjournal}

\end{document}